\documentclass[%
 amsmath,
 amssymb,
 showkeys,
 superscriptaddress,
 pra,
 aps,
 floatfix,
 10pt,
]{revtex4-2}

\usepackage{amsmath}
\usepackage{graphicx}
\usepackage{xcolor}
\usepackage{amsfonts}
\usepackage{tikz}
\usepackage{braket}
\usepackage{quantikz}
\usepackage{algorithm}
\usepackage{verbatim}
\usepackage{multirow}
\usepackage{xcolor}
\usepackage{hyperref}
\hypersetup{
    colorlinks=true,
    linktoc=all,
    linkcolor=black,
    citecolor=magenta,
}
\usepackage{titlesec}
\usepackage[title]{appendix}

\definecolor{myblue}{RGB}{0, 102, 204}
\usepackage{algorithmic}
\usetikzlibrary{decorations.pathreplacing, arrows.meta, positioning}

\usepackage{pgfplots}
\pgfplotsset{compat=1.18}
\usepackage{booktabs}
\usepackage[version=4]{mhchem}
\usepackage{verbatim}

\usepackage{caption}
\usepackage{subcaption}
\DeclareUnicodeCharacter{2212}{\textminus}

\begin{document}

\title{Quantum compilation framework for data loading}


\author{Guillermo Alonso-Linaje}
\thanks{These authors contributed equally to this work.\\ Emails: \href{mailto:guillermo@xanadu.ai}{guillermo@xanadu.ai}, \href{mailto:utkarsh@xanadu.ai}{utkarsh@xanadu.ai}.}
\author{Utkarsh Azad}
\thanks{These authors contributed equally to this work.\\ Emails: \href{mailto:guillermo@xanadu.ai}{guillermo@xanadu.ai}, \href{mailto:utkarsh@xanadu.ai}{utkarsh@xanadu.ai}.}
\affiliation{Xanadu, Toronto, ON, M5G 2C8, Canada}

\author{Jay Soni}
\affiliation{Xanadu, Toronto, ON, M5G 2C8, Canada}
\author{\\Jarrett Smalley}
\affiliation{Rolls-Royce plc., P.O. Box 31, Derby, DE24 8BJ, United Kingdom}
\author{Leigh Lapworth}
\affiliation{Rolls-Royce plc., P.O. Box 31, Derby, DE24 8BJ, United Kingdom}
\author{Juan Miguel Arrazola}
\affiliation{Xanadu, Toronto, ON, M5G 2C8, Canada}

\begin{abstract}
Efficient encoding of classical data into quantum circuits is a critical challenge that directly impacts the scalability of quantum algorithms. In this work, we present an automated compilation framework for resource-aware quantum data loading tailored to a given input vector and target error tolerance. By explicitly exploiting the trade-off between exact and approximate state preparation, our approach systematically partitions the total error budget between precision and approximation errors, thereby minimizing quantum resource costs. The framework supports a comprehensive suite of state-of-the-art methods, including multiplexer-based loaders, quantum read-only memory (QROM) constructions, sparse encodings, matrix product states (MPS), Fourier series loaders (FSL), and Walsh transform-based diagonal operators. We demonstrate the effectiveness of our framework across several applications, where it consistently uncovers non-obvious, resource-efficient strategies enabled by controlled approximation. In particular, we analyze a computational fluid dynamics workflow where the automated selection of MPS state preparation and Walsh transform-based encoding, combined with a novel Walsh-based measurement technique, leads to resource reductions of over four orders of magnitude compared to previous approaches. We also introduce two independent advances developed through the framework: a more efficient circuit for $d$-diagonal matrices, and an optimized block encoding for kinetic energy operators. Our results underscore the indispensable role of automated, approximation-aware compilation in making large-scale quantum algorithms feasible on resource-constrained hardware.

\end{abstract}

\maketitle

\setlength{\parskip}{0.15em}

\section{Introduction}

The ability to load data efficiently into quantum circuits is a fundamental prerequisite for unlocking the potential of quantum algorithms. Data loading is an important component in quantum algorithms for quantum chemistry~\cite{Aspuru_Guzik_2005, Bauer_2020, ge2018fastergroundstatepreparation}, quantum machine learning~\cite{Biamonte_2017, Huang_2021}, differential equations~\cite{shaviner2025, Zanger_2021, Krovi_2023} and computational fluid dynamics~\cite{lapworth2024evaluatio, zhuang2025}. In these diverse settings, the overall performance, fidelity, and scalability of the computation depend critically on the preparation of specific quantum states, or on the encoding of classical vectors into either the amplitudes of a quantum register or the diagonal elements of a unitary operator~\cite{cortese2018, PRXQuantum.6.020319}. This must be achieved with high fidelity and minimal resource overhead. Although a rich variety of algorithms for data loading have been proposed~\cite{2024arXiv240511436L, Araujo_2021, fomichev2024initial, gosset2024quantum}, selecting the most resource-efficient strategy for a given input vector, target accuracy, and application remains a problem-dependent task with no unique solution. \\

In this work, we introduce an automated compilation framework to navigate the trade-offs inherent in quantum data loading. Given an input vector and a user-defined error tolerance $\varepsilon$, we systematically evaluate multiple data-loading strategies and identify the optimal method requiring the fewest resources~\cite{rrreqvl-github}. At its core, the framework integrates a dedicated and extensible resource-estimation workflow built upon the PennyLane quantum software library~\cite{2018arXiv181104968B}. By automating this process, our approach algorithmically uncovers resource-efficient compilation pathways and novel implementation strategies that are often missed by static, manually designed pipelines. A key element of our approach is the explicit inclusion of approximation errors in data loading; rather than always demanding that an input vector be loaded exactly on the quantum computer, we incorporate a tolerable approximation error that allows navigation of accuracy–cost trade-offs. Across several applications, preparing approximate states yields significant cost reductions compared to exact methods. The complete code implementing the quantum compilation framework is available in the following \href{https://github.com/XanaduAI/QCFDL}{GitHub repository}. \\

Benchmarking experiments underscore the usefulness of this automated approach. For example, our framework discovered that certain Gaussian states can be implemented more efficiently by applying a quantum Fourier transform to a related, but distinct, Gaussian profile that is trivial to generate in the computational basis. This insight, which leverages global transformations to simplify local state preparation, was discovered automatically by our system without prior human input and has been corroborated in related work~\cite{xie2025efficient}. Similarly, in simulations for computational fluid dynamics (CFD) representing smooth, continuous-like velocity fields, our framework consistently found that matrix product state (MPS) methods~\cite{berry2024rapid, martin2024comb, fomichev2024initial} achieved significantly higher fidelity than competing approaches. \\

Our study also introduces three additional advances developed independently and validated through the use of our framework. First, we present a new quantum circuit for block-encoding $d$-diagonal matrices, which consolidates all non-zero diagonals into a single enlarged diagonal within a larger matrix, thereby standardizing the encoding structure and reducing the total implementation cost. Second, we propose a more efficient block-encoding construction for kinetic energy operators, which are ubiquitous in Hamiltonian simulation. Third, we demonstrate the application of Walsh-Hadamard transforms to significantly reduce the number of measurement shots required to achieve a target precision in selected CFD simulations, effectively lowering the classical overhead of the computation. \\

The remainder of this paper is organized as follows: Sec.~\ref{sec:error_model} introduces the architecture of the compilation framework and details its core components for state preparation and diagonal operator encoding. Sec.~\ref{sec:be} describes novel methodologies developed in this work for block-encoding $d$-diagonal matrices and kinetic energy operators. In Sec.~\ref{sec:applications} we present extensive numerical experiments to validate our approach, including a detailed case study of a CFD simulation, and provide a comparative analysis of resource requirements against previous unstructured, manual approaches. Finally, Sec.~\ref{sec:conclusion} summarizes the main contributions and outlines potential extensions of the framework to broader classes of quantum algorithms.

\section{Automated compilation framework}
\label{sec:error_model}

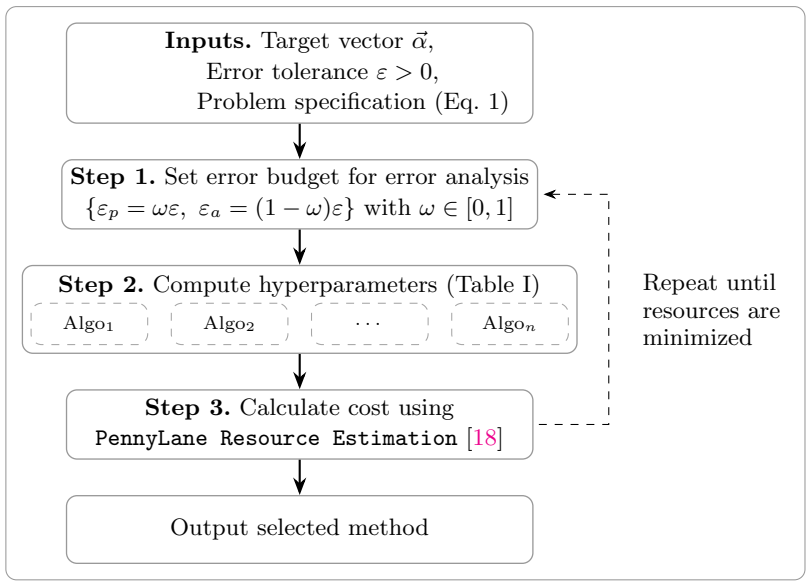
\begin{figure}[t]
\centering
\begin{tikzpicture}[
  node distance=4.7mm,
  arrow/.style={Stealth-, thick, shorten >=2pt, shorten <=2pt}, 
  block/.style={
    rectangle, rounded corners, draw=black!40, line width=0.5pt,
    minimum width=6.2cm, minimum height=9mm, align=center, font=\small
  },
  elbow arrow/.style={
    dashed,
    -{Stealth},
    shorten >=2pt,
    shorten <=2pt,
    to path={(\tikztostart) -- ++(#1,0) |- (\tikztotarget) \tikztonodes}
  },
  subblock/.style={
    rectangle, draw=black!35, dashed,
    minimum width=1.55cm, minimum height=5.5mm,
    align=center, font=\scriptsize
  },
  label/.style={font=\small, align=left}
]
\node[block] (step1) {\textbf{Inputs.}~Target vector $\vec{\alpha}$,\\[1pt]\quad\;\; Error tolerance $\varepsilon > 0$,\\[1pt]\quad\;\quad\;\quad\;\;\;Problem specification {(Eq.~\ref{eq:prob-type})}};

\node[block, below=of step1] (step2) {\textbf{Step 1.}~Set error budget for error analysis\\[1pt]
\(\{\varepsilon_p=\omega\varepsilon,\ \varepsilon_a=(1-\omega)\varepsilon\}\) with $\omega\in[0,1]$};
\node[block, below=of step2] (step3) {%
  \textbf{Step 2.}~Compute hyperparameters (Table~\ref{tab:combined_errors})\\[1pt]
  \begin{tikzpicture}[every node/.style={subblock}]
\node (alg1) {Algo$_1$};
\node[right=3mm of alg1] (alg2) {Algo$_2$};
\node[right=3mm of alg2] (dots) {$\dots$}; 
\node[right=3mm of dots] (algn) {Algo$_n$};
  \end{tikzpicture}
};
\node[block, below=of step3] (step4) {\textbf{Step 3.}~Calculate cost using \\[1pt] \texttt{PennyLane Resource Estimation} \cite{2018arXiv181104968B}};
\node[block, below=of step4] (keep) {Output selected method};

\coordinate (annX) at ([xshift=.75cm]step3.east);
\node[label, anchor=west] (labOPT) at (annX |- step3.east) {Repeat until \\resources are\\minimized};

\node[
  draw=black!40, rounded corners, fit=(step1)(step2)(step3)(step4)(labOPT)(keep),
  inner sep=6pt, name=container
] {};


\draw[-Stealth, thick] (step1) -- (step2);
\draw[-Stealth, thick] (step2) -- (step3);
\draw[-Stealth, thick] (step3) -- (step4);
\draw[-Stealth, thick] (step4) -- (keep);
\draw[elbow arrow] (step4.east) to (step2.east);



\end{tikzpicture}
\caption{\textbf{Workflow of the automated compilation framework}. Given a vector $\vec{\alpha}$ and an error tolerance $\varepsilon$, a weight $w$ is selected by taking different values of a grid partition and distributing the total error between the precision error and the approximation error.  
Once $w$ is fixed, the optimal hyperparameters of the different algorithms are calculated, and the resources are estimated using the PennyLane resource estimation framework.  
The value of $w$ is then updated, and finally, we return the most efficient method found previously.}
\label{fig:diagram}
\end{figure}

We address the challenge of loading classical data onto a quantum computer by tackling two fundamental problems: \textit{state preparation}~\cite{fomichev2024initial, grover2002creating} and \textit{diagonal block encoding}~\cite{camps2023explicit, zylberman2025efficient}, each with a distinct workflow. Although their objectives differ, they share an underlying design philosophy: automating the selection and configuration of algorithms to meet a prescribed accuracy $\varepsilon$ with minimal quantum resources. By treating these tasks within a single cohesive system, we can systematically compare disparate methods on equal footing. \\

Formally, the framework aims to identify an optimal unitary approximation, denoted by \(\hat{U}'\), of a target unitary operator \(\hat{U}\), such that it satisfies the prescribed error bounds for the two considered problems. Let \(\vec{\alpha} = (\alpha_0, \ldots, \alpha_{d-1}) \in \mathbb{C}^d\) denote the target vector to be loaded on the quantum computer. Then the target problems are defined in terms of finding a circuit implementing the unitary \(\hat{U}'\) such that for a target approximation error $\varepsilon$ the following holds:
\begin{equation}
\label{eq:prob-type}
\begin{split}
(a) &\ \ \textbf{State Preparation:} 
\quad \langle i | \hat{U} | 0 \rangle = \alpha_i
\ \ \Longrightarrow \ \ 
\langle i | \hat{U}' | 0 \rangle = \alpha_i', 
\quad \text{with} \quad  
\| \vec{\alpha} - \vec{\alpha}' \|_2 \leq \varepsilon, \\[6pt]
(b) &\ \ \textbf{Diagonal Encoding:} 
\quad \langle i | \hat{U} | i \rangle = \alpha_i
\ \ \Longrightarrow \ \ 
\langle i | \hat{U}' | i \rangle = \alpha_i', 
\quad \text{with} \quad  
\| \hat{U} - \hat{U}' \|_2 \leq \varepsilon,
\end{split}
\end{equation}
where \(\|\cdot\|_2\) denotes the \(\ell_2\)-norm, and in the diagonal case it is associated with the block-encoded matrix instead of the total matrix operator.\\

The ultimate goal is to replace manual, heuristic-based design with an automatic optimization process that is both rigorous and adaptable. To achieve this, we have designed a systematic workflow, illustrated in Fig.~\ref{fig:diagram}, which methodically navigates the trade-off between accuracy and implementation cost. A key feature of this process is its explicit handling of the total error budget, $\varepsilon$. Rather than treating error as a monolithic quantity, the framework begins by partitioning it into two distinct error categories, balanced by a weighting parameter $\omega \in (0, 1]$: (i) an approximation error $\varepsilon_a = (1 - \omega)\,\varepsilon \geq 0$, which is deliberately introduced to allow potential reductions in circuit depth or gate count compared to exact data loading, and (ii) precision error $\varepsilon_p = \omega\,\varepsilon> 0$, which arises unavoidably from the finite precision of gate operations in specifying rotation angles and synthesizing high-level operations into native gates. This partitioning acts as a strategic knob, allowing us to explore whether it is cheaper to implement a simplified problem perfectly or an exact problem imperfectly, a crucial decision that deeply impacts the final resource count. A grid search is used to determine the best partition. With a specific error budget $(\varepsilon_a, \varepsilon_p)$ established, the framework then explores a diverse portfolio of candidate algorithms, which we review in the appendices \hyperref[sec:stateprep]{A}-\hyperref[sec:diag]{B} and summarize in the Table \ref{tab:combined_errors}.\\


\begin{table}[t]
\centering
\begin{tabular}{@{}p{1.0cm}p{4.5cm}p{4.5cm}p{4.3cm}p{3cm}@{}}
\toprule
\textbf{Task} & \textbf{Method} & \textbf{Precision error} ($\varepsilon_p$) & \textbf{Approximation error} ($\varepsilon_a$) & \textbf{Hyperparameters} \\
\midrule \\ 
\multirow{6}{*}{\rotatebox[origin=c]{90}{\begin{tabular}{l}\textbf{State}\\\textbf{Preparation}\end{tabular}}}
& Möttönen~\cite{mottonen2004transformation} & Rotation synthesis & None & $\delta_G$ [Eq.~\ref{eq:hyp-mot}]\\
& QROM state preparation~\cite{grover2002creating} & Truncation of rotation angles & None & $m$ [Eq.~\ref{eq:hyp-qrom-sp}]\\
& Sparse~\cite{fomichev2024initial} & QROM state preparation error & Use top $D$ amplitudes & $(m, D)$\\
& Matrix product state~\cite{fomichev2024initial} & Rotation synthesis & MPS compression & ($\delta_G,\ \chi$) [Eq.~\ref{eq:hyp-mps}]\\
& Fourier Series Loader~\cite{Moosa_2023} & QROM state preparation error & Use $d$ Fourier coefficients & $(m, d)$\\
& Alias sampling~\cite{Babbush_2018} &  Truncation of amplitudes & None & $\mu$ [Eq.~\ref{eq:hyp-alias}]\\ \\
\midrule \\
\multirow{3}{*}{\rotatebox[origin=c]{90}{\begin{tabular}{l}\textbf{Diagonal}\\\textbf{Encoding}\end{tabular}}}
& Möttönen~\cite{mottonen2004transformation} & Rotation synthesis & None & $\delta_G$ [Eq.~\ref{eq:hyp-mot2}]\\
& QROM~\cite{grover2002creating} & Truncation of rotation angles & None & $m$\\
& Quantum signal processing~\cite{mcardle2025quantum} & Rotation synthesis & Polynomial approximation & $\delta_G$ [Eq.~\ref{eq:hyp-mot}]\\
& Walsh transform~\cite{zylberman2025efficient} & Rotation synthesis & Walsh approximation & ($\delta_G$, $\kappa$) [Eq.~\ref{eq:hyp-walsh}] \\[10pt]
\bottomrule
\end{tabular}
\caption{\textbf{Sources of precision $\varepsilon_p$ and approximation $\varepsilon_a$ errors for data-loading methods.} Here, the \textit{truncation} for precision errors refers to truncating the binary representation of the specified values to $m$ (or $\mu$) bits. In addition to this, the QROM state preparation error implies the precision error arising from the underlying state preparation procedure used in the method, as discussed in the Appendix \hyperref[sec:stateprep]{A}. Moreover, the $\delta_G \geq 0$ is the per-gate rotation synthesis error, $\chi$ is the bond dimension that relates to the compressibility of the MPS, and $d, D, \kappa > 0$ refers to the truncation order specifying the number of terms used for approximation.}
\label{tab:combined_errors}
\vspace{-8pt}
\end{table}

 Each state preparation and diagonal block encoding algorithm that we consider is accompanied by its primary error sources, parameterized by a set of tunable hyperparameters, such as discretization granularity, decomposition depth, or approximation rank. We also present a detailed error analysis that determines a feasible parameter space that satisfies the error budget. The resulting configurations are then benchmarked using the integrated resource-estimation workflow, allowing the framework to select the most resource-efficient solution. For example, a larger approximation error $\varepsilon_a$ might permit a lower bond dimension for an MPS, or fewer coefficients for a Fourier series loader (FSL), potentially leading to a shallower circuit. This step effectively translates the abstract error budget into a concrete set of valid implementation strategies for every available method, preparing them for a final head-to-head comparison.\\

To guide hyperparameter optimization, we measure synthesis error using the $\ell_2$-norm of individual gate errors, while noting that other approaches, such as cumulative-error models, are also common in the literature \cite{ma2021markovchainshittingtimes}. It is therefore useful to keep track of the number of rotation gates, since each rotation contributes to the synthesis error when compiled into a gate set such as Clifford+T \cite{Kliuchnikov_2023}. Once the hyperparameters are fixed, we also rely on an efficient and scalable method for estimating the resource requirements of each technique. For this purpose, we employ PennyLane’s resource estimation functionality, which provides accurate estimates without explicit construction of large circuits.\\

\begin{figure}[!t]
\centering
\begin{tikzpicture}[baseline=(current bounding box.center), scale=1., every node/.style={scale=1.}]

\node (circuit) at (7.5, 0) {
\begin{quantikz}[row sep=0.5cm, column sep=0.6cm]
\lstick{} & \octrl{1} & \octrl{1} & \ctrl{2} & \qw \\
\lstick{} & \octrl{1} & \ctrl{1} & \qw & \qw \\
\lstick{} & 
    \gate[style={fill=blue!20}, wires=1]{\text{SP}_1} & 
    \gate[style={fill=red!20}, wires=1]{\text{SP}_2} & 
    \gate[style={fill=brown!30}, wires=1]{\text{SP}_3} & \qw \\
\end{quantikz}
};

\node (hist) [left=7.5cm of circuit.center, anchor=center] {
\begin{tikzpicture}[scale=0.9, every node/.style={scale=0.9}]
\begin{axis}[
    ybar,
    bar width=6pt,
    width=9cm,
    height=5cm,
    ymin=0, ymax=3.2,
    xtick=\empty,
    ytick=\empty,
    axis x line=bottom,
    axis y line=left,
    enlarge x limits=0.06,
]

\addplot+[ybar, bar shift=0pt, draw=black, fill=blue!50] coordinates {
    (1,1.2) (2,0.8) (3,1.1) (4,1.6)
};

\addplot+[ybar, bar shift=0pt, draw=black, fill=red!40] coordinates {
    (5,0.2) (6,0.2) (7,0.2) (8,0.22)
};

\addplot+[ybar, bar shift=0pt, draw=black, fill=brown!40] coordinates {
    (9,2.2) (10,2.7) (11,2.3) (12,2.8) (13,2.2) (14,2.7) (15,1.3) (16,0.8)
};
\end{axis}

\draw[decorate,decoration={brace,mirror,amplitude=5pt}, thick]
    (0.0,-0.2) -- (1.65,-0.2);
\node at (0.95,-0.7) {\small SP$_1$};

\draw[decorate,decoration={brace,mirror,amplitude=5pt}, thick]
    (1.85,-0.2) -- (3.6,-0.2);
\node at (2.75,-0.7) {\small SP$_2$};

\draw[decorate,decoration={brace,mirror,amplitude=5pt}, thick]
    (3.8,-0.2) -- (7.4,-0.2);
\node at (5.7,-0.7) {\small SP$_3$};

\end{tikzpicture}
};

\draw[->] (hist) -- node[midway, above] {} (circuit);

\end{tikzpicture}
\caption{\textbf{Example of hybrid state preparation.} The histogram (left) reveals structural patterns in the target state. Our recursive method applies different controlled routines (SP$_1$, SP$_2$, SP$_3$) in the quantum circuit (right) to efficiently prepare each region.}
\label{fig:hybrid_sp}
\end{figure}
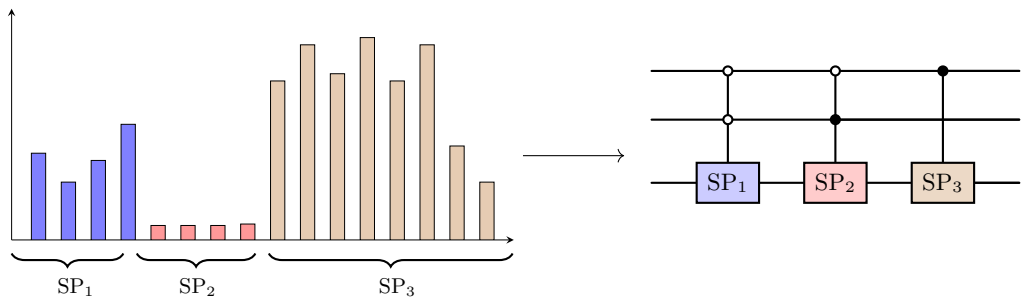

Beyond selecting a single algorithm, the framework also supports \emph{hybridization} of approaches \cite{Gui_2024}, allowing the preparation of a target state using multiple algorithms applied to different segments of the input vector. This is implemented through controlled versions of the individual routines, where a register of control qubits selects the subspace in which each partial state is prepared. This divide-and-conquer approach allows us to partition a data vector into several regions and apply the most suitable algorithm to each, controlled via ancillary qubits. This is especially effective for complex data with varying local structures; for instance, a sparse region might be handled by one method and a smooth, wavelike section by another, all within a single quantum circuit. For example, Fig.~\ref{fig:hybrid_sp} illustrates a state being partitioned into three intervals, with each assigned to a distinct preparation routine (e.g., SP\(_1\), SP\(_2\), SP\(_3\)). The partitions are enforced using binary control logic, which requires that each subvector correspond to a contiguous block of the Hilbert space aligned with binary index boundaries. \\

In summary, the automated compilation framework provides a unified methodology for selecting, parameterizing, and combining data loading algorithms while rigorously accounting for both approximation and precision errors. Explicitly incorporating approximation errors and integrating them with scalable resource estimation enables systematic navigation of algorithmic design choices that can be tailored to reduce the implementation cost for a wide range of quantum applications. In the following sections, we describe new results developed using our framework, then illustrate the usefulness of the automated compilation scheme across a variety of applications.\\

\section{Block-encoding methods}
\label{sec:be}
In this section, we describe novel methodologies developed in this work for block-encoding $d$-diagonal matrices and block-encoding kinetic energy operators, extending the applicability of the compilation framework to more complex scenarios.

\subsection{Encoding of d-diagonal matrices}

The $d$-diagonal matrices where all entries are zero except for those lying on $d$ specified diagonals, arise frequently in computational problems \cite{lapworth2024evaluatio}. For instance, such matrices appear naturally in the discretization of differential equations after linearizing the underlying problem \cite{lapworth2025precondition}. A common approach for constructing a block encoding of a $d$-diagonal matrix involves a linear combination of unitaries (LCU) circuit, which encodes each diagonal individually. Typically, this is done by first constructing a block encoding of the primary diagonal, i.e., a 1-diagonal block encoding, and then shifting it to its correct position using an arithmetic adder. This technique has been employed in prior works such as~\cite{lapworth2025precondition}, where the number of arithmetic operators grows linearly with the number of diagonals.\\

\begin{figure*}[!t]
    \centering
    \begin{subfigure}[b]{.46\linewidth}
    \begin{minipage}{.03\textwidth}
        \caption{}
        \label{fig:diags2}
    \end{minipage}%
    \begin{minipage}{0.95\textwidth}
      \scalebox{0.98}{
        \begin{quantikz}[row sep=1.2em, column sep=0.4em]
          \lstick{}
            & \qw 
            & \qw 
            & \qw 
            & \qw 
            & \gate[wires=2]{D_k}
            & \qw 
            & \qw
          \\
          \lstick{}
            & \qw
            & \qw
            & \gate[wires=2]{Adder}
            & \qw
            & \ghost{D_k}
            & \qw
            & \qw
          \\
          \lstick{}
            & \qw
            & \gate{Load\,k}
            & \ghost{Adder}
            & \gate{Load\,k^\dagger}
            & \ghost{D_k}
            & \qw
            & \qw
          \\
          \lstick{}
            & \gate{Prep}
            & \ctrl[style={draw,fill=white,shape=rectangle}]{-1}
            & \qw
            & \ctrl[style={draw,fill=white,shape=rectangle}]{-1}
            & \ctrl[style={draw,fill=white,shape=rectangle}]{-2}
            & \gate{Prep^{\dagger}}
            & \qw\\
        \end{quantikz}
        
      }
    \end{minipage}
    \end{subfigure}
    \begin{subfigure}[b]{.52\linewidth}
    \begin{minipage}{.05\textwidth}
        \caption{}
        \label{fig:diagonal_approx}
    \end{minipage}%
    \begin{minipage}{0.95\textwidth}
        \includegraphics[width=\linewidth]{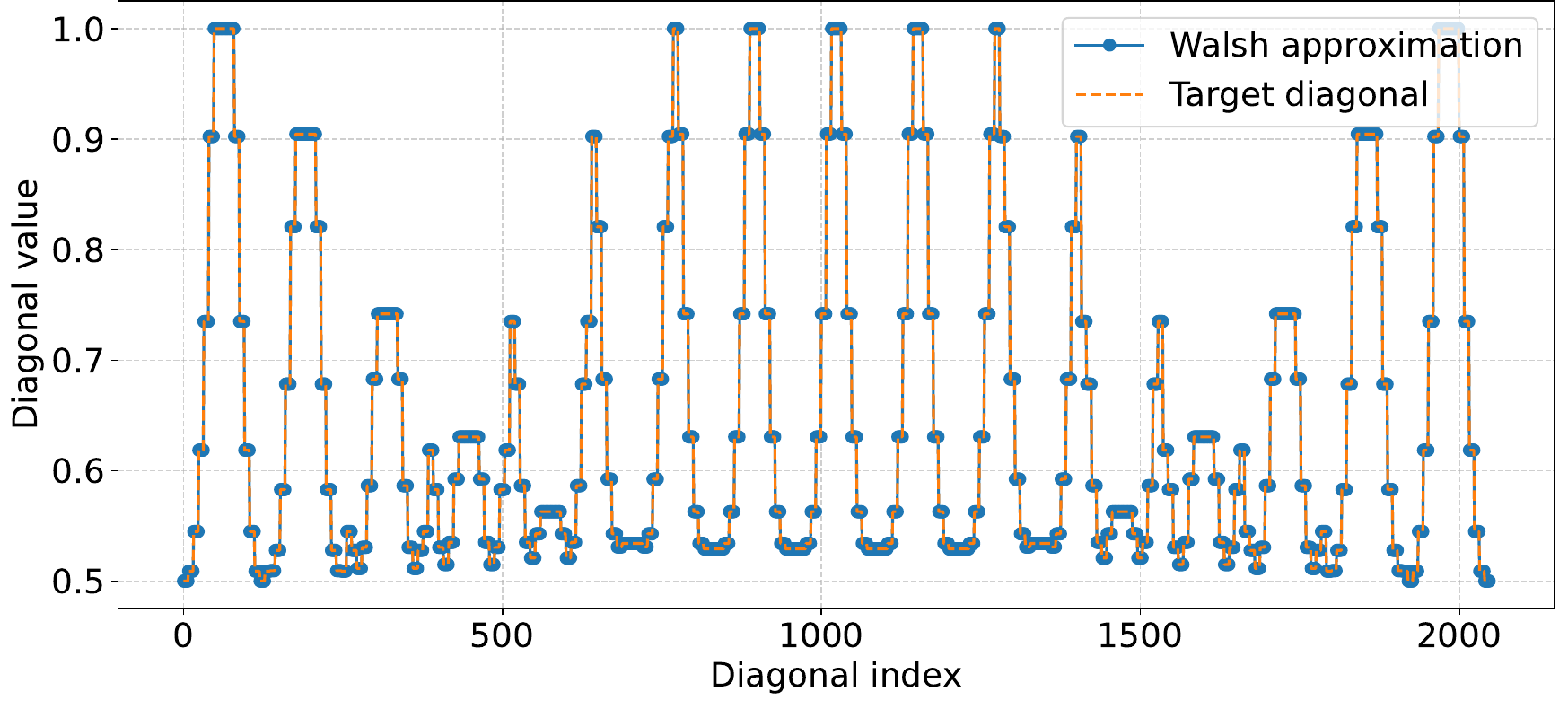}
    \end{minipage}
    \end{subfigure}
    \caption{\textbf{Block encoding of d-diagonal matrices.} (a) Compact restructuring of the diagonal block encoding by unifying the quantum in-place Adders into a single Adder. Here we also use $D_k$ as the block encoding of the $k$-th diagonal. (b) Reconstruction using the 64 most significant Walsh coefficients, where the x-axis and y-axis show the diagonal's index and value, respectively.}
    \label{fig:d-diag-encoding}
\end{figure*}

We present an improved method that reduces the arithmetic overhead to a single operator. Our approach enables the simultaneous application of the $d$ shifts in quantum superposition, effectively simulating the combined effect of all $d$ adders with a single arithmetic operation. It is illustrated in Fig. \ref{fig:diags2}, where we employ square-control notation to denote multiplexed operations. 
The target block encoding is defined as \(\sum_{i = 1}^d \alpha_i D_i\), where \(D_i\) denotes a diagonal matrix shifted by \(k_i\), expressed as 
\(D_i := \sum_{j} c_{ij}\, |j + k_i\rangle\langle j|\),
with \(c_{ij}\) representing the diagonal entries. The algorithm proceeds as follows:

\begin{enumerate}
    \item \textbf{Initial state preparation.} Prepare the superposition state \(\text{Prep}\ket{0} = \sum_i \sqrt{\alpha_i}\ket{i}\), where the coefficients \(\alpha_i\) weight the different diagonals.
    
    \item \textbf{Loading the diagonal shifts.} Encode the shift values \(k_i\) through multiplexed operations, yielding \(\sum_i \sqrt{\alpha_i}\ket{i}\ket{k_i}\).
    
    \item \textbf{In‐place addition in superposition.} Apply the addition operation in superposition to obtain 
    \begin{equation}
    \sum_i \sqrt{\alpha_i}\ket{k_i}\ket{i}\mathrm{Adder}(k_i)
    = \sum_i \sqrt{\alpha_i}\ket{k_i}\ket{i}\!\bigg[\sum_j \ket{j + k_i}\bra{j}\bigg].        
    \end{equation}
    
    This step uses a half adder \(\mathrm{Adder}(\cdot)\) that adds the constant \(k_i\) to the second register. A single application of this adder generates a coherent superposition over all \(d\) in-place adders that would otherwise need to be applied sequentially.

    \item \textbf{Uncomputation and diagonal encoding.} Apply the adjoint of the loading gate to uncompute \(\ket{k_i}\), followed by the diagonal block encoding, resulting in block-encoding of the operator
    \begin{equation}
    \sum_i \sqrt{\alpha_i}\ket{i}\!\bigg[\sum_j c_{ij}\ket{j + k_i}\bra{j}\bigg] = \sum_i \sqrt{\alpha_i}\ket{i}D_i.
    \end{equation}
    
    \item \textbf{Linear combination of weighted diagonals.} Finally, apply \(\mathrm{Prep}^\dagger\) to coherently combine the weighted diagonals, recovering the operator
    \(\sum_{i = 1}^d \alpha_i D_i\) on the top-left block of the overall encoded unitary.
\end{enumerate}

While the given construction reduces the overall cost of the algorithm, the bottleneck remains in synthesizing \(D_k\). However, a significant advantage of our circuit is that, unlike previous approaches \cite{lapworth2025precondition}, 
it enables the unification of all diagonals into a single, larger diagonal. This feature allows the simultaneous approximation of all diagonals 
using a single algorithm. 

\subsection{Block encoding for kinetic energy operators}

An efficient block encoding of the kinetic energy operator can significantly reduce the resources required for quantum simulation workflows. As described in~\cite{Su_2021}, in the first quantization, the kinetic energy operator $\hat{T}$ is defined as:
\begin{align}
    \hat{T} 
    &= \sum_{j=1}^{\eta} \sum_{p \in G} 
    \frac{\|k_p\|^2}{2} \ket{p}\bra{p}_j \nonumber \\[4pt]
    &= \sum_{x,y,z} 
    \frac{1}{2} \left( \frac{2\pi}{\Omega^{1/3}} \right)^2 
    (x^2 + y^2 + z^2)\ket{xyz}\bra{xyz} \nonumber \\[6pt]
    &= 
    \sum_{x,y,z} \frac{1}{2} \left( \frac{2\pi}{\Omega^{1/3}} \right)^2 
    \big( x^2\ket{xyz}\bra{xyz} 
    + y^2\ket{xyz}\bra{xyz} 
    + z^2\ket{xyz}\bra{xyz} \big) \nonumber \\[6pt]
    &= 
    \frac{1}{2} \left( \frac{2\pi}{\Omega^{1/3}} \right)^2 
    \Big[ 
      \big( \sum_x x^2 \ket{x}\bra{x} \big) \!\otimes\! \mathbb{I} \!\otimes\! \mathbb{I}
      + \mathbb{I} \!\otimes\! \big( \sum_y y^2 \ket{y}\bra{y} \big) \!\otimes\! \mathbb{I}
      + \mathbb{I} \!\otimes\! \mathbb{I} \!\otimes\! \big( \sum_z z^2 \ket{z}\bra{z} \big)
    \Big] \nonumber \\[6pt]
    &=: 
    \hat{T}_x \!\otimes\! \mathbb{I} \!\otimes\! \mathbb{I}
    + \mathbb{I} \!\otimes\! \hat{T}_y \!\otimes\! \mathbb{I}
    + \mathbb{I} \!\otimes\! \mathbb{I} \!\otimes\! \hat{T}_z.
\end{align}
where \( (x,y,z) \) discretize the spatial grid. The key advantage of this representation is that the operator could be decomposed into the sum of three separate block encodings ($\hat{T}_{x/y/z}$). Each of these block encodings is 1-diagonal, and they can be exactly implemented ($\varepsilon_a = 0$) via a degree-2 polynomial. This means that the diagonal operator can be encoded exactly using a quantum circuit composed of polynomial transformations of degree two using a quantum signal processing approach as described in the Appendix \hyperref[sec:QSP]{B.2}.\\

\begin{figure}[t]
    \centering
\begin{quantikz}[row sep=0.05cm, column sep=0.5cm]
\lstick{$\ket{0}$} & \gate[wires=2]{\text{Prep}} & \octrl{1} & \octrl{2} & \ctrl{3} & \gate[wires=2]{\text{Prep}^\dagger} & \qw \\
\lstick{$\ket{0}$} &                               & \octrl{1} & \ctrl{2} & \octrl{3} &                              & \qw \\
\lstick{$x$} & \qw & \gate{\hat{T}_x^{(2)}} & \qw & \qw & \qw & \qw \\
\lstick{$y$} & \qw & \qw & \gate{\hat{T}_y^{(2)}} & \qw              & \qw & \qw \\
\lstick{$z$} & \qw & \qw & \qw              & \gate{\hat{T}_z^{(2)}} & \qw & \qw
\end{quantikz}

\caption{\textbf{Block encoding circuit for the kinetic operator.} The operators $\hat{T}_x$, $\hat{T}_y$, and $\hat{T}_z$ are prepared using a quantum signal processing approach as described in the Appendix
\hyperref[sec:QSP]{B.2}.}
    \label{fig:kinetic}
\end{figure}
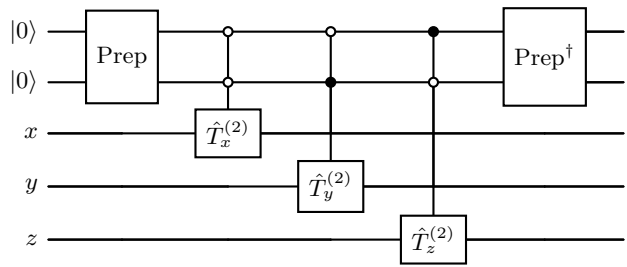

Notably, our framework was able to automatically detect that with $d=2$, this technique was actually the most resource-efficient. This contrasts with other methods that build an arithmetic square operator \cite{Su_2021}, achieving similar results but at the cost of an increased number of qubits.
The corresponding circuit is illustrated in Fig.~\ref{fig:kinetic}, where the LCU-style blocks represent the diagonal block encodings of the three operators defined above.

\section{Applications}
\label{sec:applications}

In this section, we demonstrate the applicability of our framework in three distinct scenarios: (i) the preparation of Gaussian states, (ii) the ground-state preparation of different simple molecules, and (iii) its integration into a complete workflow for fluid dynamics simulation. For each case, we present the selected method, compare the required resources against alternative techniques, and evaluate the relative weight between these two sources of error. These results reveal that, in comparison with exact approaches, the required resources can be automatically reduced by orders of magnitude.

\begin{figure}[!thp]
    \centering
    \begin{subfigure}[b]{0.60\textwidth}
        \centering
            \includegraphics[width=\textwidth]{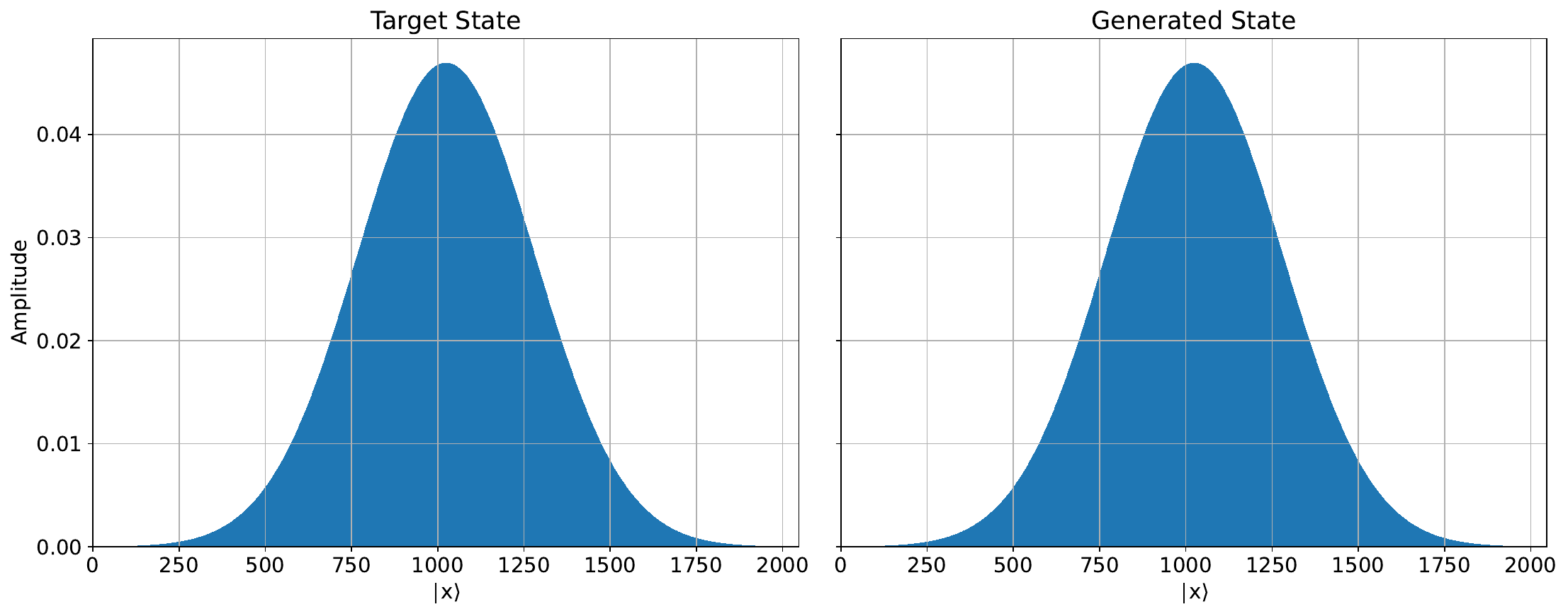}
        \caption{State Reconstruction with FSL}
        \label{fig:gaus}
    \end{subfigure}
    \begin{subfigure}[b]{0.39\textwidth}
        \centering
            \includegraphics[width=\textwidth]{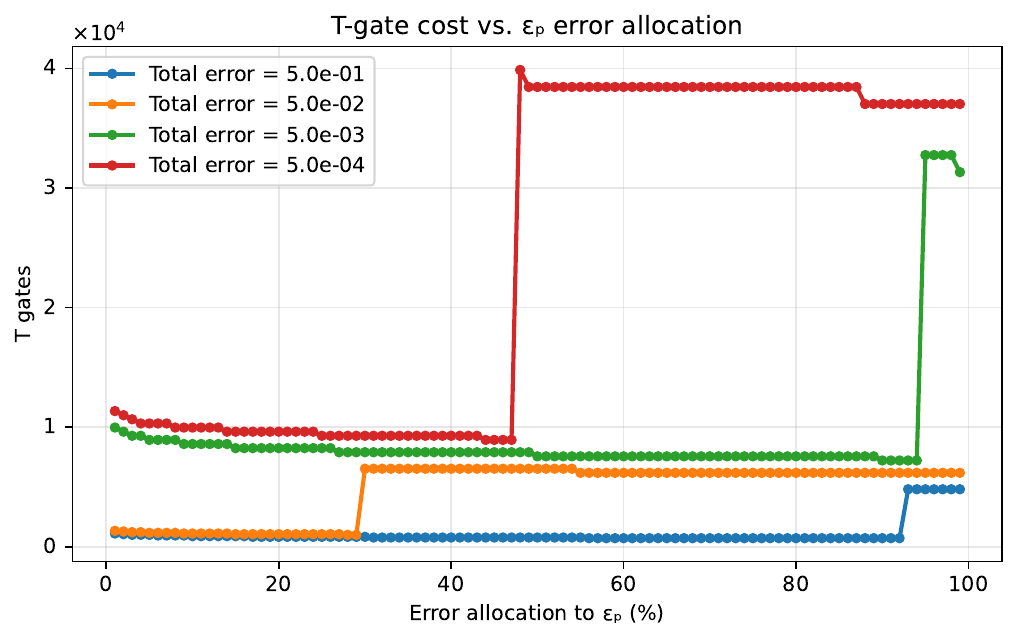}
        \caption{T-gate cost vs. error allocation ($\omega$)}
        \label{fig:error_allocation}
    \end{subfigure}
    \caption{\textbf{State preparation of Gaussian states using FSL.} (a) Efficient preparation of a discretized Gaussian state over 11 qubits using the Fourier Series Loader (FSL). Only 8 Fourier coefficients are required to achieve an approximation error of \(\varepsilon = 1\mathrm{e}{-4}\). (b) Trade-off for a 14-qubit Gaussian state ($\sigma=0.9$). The $x$-axis shows the fraction of the total error budget $\varepsilon$ assigned to the preparation error, $w=\varepsilon_p/\varepsilon$ (displayed in \%). The $y$-axis reports the estimated T-gate count. Each curve corresponds to a different total error $\varepsilon \in \{5\mathrm{e}{-1}, 5\mathrm{e}{-2}, 5\mathrm{e}{-3}, 5\mathrm{e}{-4}\}$. For a fixed $\varepsilon$, increasing the allocation to $\varepsilon_p$ (thereby decreasing the approximation error $\varepsilon_a=(1-w)\varepsilon$) illustrates the trade-off between preparation and approximation resources.}
    \label{fig:fsl-state-prep-gaus}
\end{figure}

\subsection{Gaussian states}

The preparation of Gaussian states arises in various quantum algorithmic contexts, particularly when working in real-space encodings or shaping energy distributions. Prominent examples include vibrational simulations \cite{McArdle_2019}, where the eigenfunctions of the harmonic oscillator often take the form of Hermite polynomials multiplied by a Gaussian envelope. Another use case appears in quantum phase estimation \cite{loaiza2025nonlinear}, where Gaussian lineshapes have been proposed as windowing functions to suppress spectral leakage. Although alternatives like the Kaiser window \cite{greenaway2024case} may offer superior asymptotic performance, the Gaussian remains a practical choice. To test the framework's ability to handle such states, we consider the preparation of a discretized Gaussian state over $11$ qubits, with zero mean and standard deviation $\sigma=0.5$, targeting an approximation error of $\varepsilon=1\mathrm{e}{-3}$. The Table \ref{tab:combined_gate_counts} (column 2) presents the resource costs for this task using different methods, demonstrating the advantage of using the Fourier series loader for such states. \\

In this instance, the framework evaluates different error distributions, ultimately determining that a weighted error of $60\%$ for the precision error $\varepsilon_p$ and 40\% for the approximation error $\varepsilon_a$ provides an optimal balance. This error allocation parameter plays a crucial role in resource estimation, as illustrated in Fig.~\ref{fig:error_allocation}, where different error allocations can result in different behaviours. The resulting state approximation based on these parameters, shown in Fig.~\ref{fig:gaus}, requires only 32 Fourier coefficients to meet the specified accuracy. \\

Given that the Fourier transform of a Gaussian is itself a Gaussian, one might expect Fourier-based methods to perform poorly due to the lack of sparsity in the frequency domain. However, because the standard deviation of the Gaussian narrows under the transformation \cite{xie2025efficient}, the Fourier profile becomes more concentrated, enabling a sparse and efficient representation. The framework automatically leverages this structure, identifying non-obvious strategies that minimize resource utilization. For example, when the total error is increased, the framework shifts its recommendation from the Fourier Series Loader to an MPS-based approach. This highlights a broader principle: even within a single functional family, small changes in parameters can drastically alter the most efficient encoding strategy, underscoring the necessity of automated and data-driven method selection for practical quantum algorithm design.

\subsection{Quantum Chemistry}

The efficient preparation of quantum states with non-trivial overlap with the ground state is a crucial requirement for quantum algorithms in electronic structure problems. In many instances, the ground state of a molecular Hamiltonian can be well-approximated by a linear combination of a small number of Slater determinants with dominant amplitudes. Identifying and exploiting this inherent sparsity enables a substantial reduction in state preparation costs. We consider this challenge in the context of the BeH$_2$ molecule from the Pennylane molecules dataset \cite{Utkarsh2023Chemistry} using the STO-3G basis set and Be$-$H bond length of $1.33$~\AA. \\

For this problem, the framework selects the sparse state preparation strategy, which constructs the quantum state by explicitly initializing only the significant amplitudes \cite{fomichev2024initial}. This approach assumes a classical description of the dominant components. It yields a circuit complexity that scales with the number of non-zero amplitudes rather than the full Hilbert space dimension, making it highly suitable for compressed quantum states. Here, the framework adopts a balanced error model, assigning $50\%$ of the total error budget to precision error $\varepsilon_p$ and $50\%$ to approximation error $\varepsilon_p$. As shown in the Table~\ref{tab:combined_gate_counts} (column 3), the sparse preparation strategy demonstrates resource efficiency for preparing the ground state of the BeH$_2$ molecule compared to other methods. Furthermore, we confirm its robust performance by preparing the ground states of even larger molecules with the STO-3G basis set, such as C$_2$H$_4$, C$_2$, and BH$_3$, yielding T-counts of $6.0 \times 10^4$, $4.6 \times 10^5$, and $2.3 \times 10^5$, respectively. \\

\begin{table}[t]
\centering
\begin{tabular}{@{}p{3.9cm}p{2.1cm}p{2.1cm}p{2.1cm}p{2.1cm}p{2.1cm}p{2.1cm}@{}}
\toprule
\multirow{2}{*}{\textbf{Method}} & \multicolumn{2}{c}{\textbf{Gaussian state}} & \multicolumn{2}{c}{\textbf{BeH$_2$ ground state}} & \multicolumn{2}{c}{\textbf{2D-LDC state}} \\
\cmidrule(lr){2-3} \cmidrule(lr){4-5} \cmidrule(lr){6-7}
& \textbf{\# CNOTs} & \textbf{\# T gates} & \textbf{\# CNOTs} & \textbf{\# T gates} & \textbf{\# CNOTs} & \textbf{\# T gates} \\
\midrule
Fourier Series Loader \cite{Moosa_2023} 
 & \textbf{1.91} $\times$ \textbf{10$^2$} & \textbf{8.86 $\times$ 10$^3$} 
 & $3.30 \times 10^4$ & $1.13 \times 10^6$
 & $2.59 \times 10^4$ & $2.53 \times 10^4$ \\
Matrix Product State \cite{fomichev2024initial} 
 & $6.25 \times 10^2$ & $3.15 \times 10^4$
 & $4.19 \times 10^4$ & $1.59 \times 10^6$
 & \textbf{2.22 $\times$ 10$^2$} & \textbf{6.24 $\times$ 10$^3$} \\
Sum of Slaters \cite{fomichev2024initial}        
 & $2.23 \times 10^4$ & $1.89 \times 10^4$
 & \textbf{4.75 $\times$ 10$^3$} & \textbf{5.42 $\times$ 10$^3$}
 & $1.59 \times 10^4$ & $1.13 \times 10^4$ \\
Mottonen State Prep \cite{mottonen2004transformation}   
 & $4.09 \times 10^3$ & $1.35 \times 10^5$
 & $3.28 \times 10^4$ & $1.11 \times 10^6$
 & $4.09 \times 10^3$ & $1.11 \times 10^5$ \\
QROM State Prep \cite{grover2002creating}      
 & $2.23 \times 10^4$ & $1.90 \times 10^4$
 & $3.15 \times 10^5$ & $2.33 \times 10^5$
 & $1.62 \times 10^4$ & $1.15 \times 10^4$ \\
\bottomrule
\end{tabular}
\caption{\textbf{Resource comparison for quantum state preparation methods.} We analyze $\mathrm{CNOT}$ and $\mathrm{T}$ gate counts for various methods for three distinct types of state vectors: (a) Gaussian state from vibrational Hamiltonian simulations, (b) electronic ground state of a BeH$_2$ molecule, and (c) state of a two-dimensional lid-driven cavity (2D-LDC) used in CFD simulations. The best-performing methods, i.e., ones with the lowest T-gates, for each state vector are highlighted in \textbf{bold}.}
\label{tab:combined_gate_counts}
\end{table}

This consistent behaviour can be attributed to the high sparsity of the corresponding ground states, which prevents the quantum resource requirements from scaling steeply with the number of qubits.

\subsection{Computational Fluid Dynamics}

The two-dimensional lid-driven cavity flow (2D-LDC)~\cite{lapworth2022implicit} is a classical benchmark in computational fluid dynamics (CFD) for validating numerical schemes that solve the incompressible Navier–Stokes equations.
The setup consists of a square cavity with a side length of \(L\), filled with a constant-density Newtonian fluid.  
The top wall (lid) moves horizontally at a constant velocity \(U\), while the other three walls remain stationary, all satisfying no-slip boundary conditions.
The Reynolds number (Re) characterizes the flow by
\begin{equation}
    \text{Re} = \frac{\rho\;U L}{\mu},
\end{equation}
where \(\rho\) is the fluid density and \(\mu\) the dynamic viscosity. To solve the Navier–Stokes equations, we employ the SIMPLE (Semi-Implicit Method for Pressure-Linked Equations) algorithm~\cite{PATANKAR19721787}, in which the nonlinear convective terms are linearized at each iteration, producing a sequence of sparse linear systems with a fixed sparsity pattern \cite{lapworth2022hybrid}.  
In particular, the discretized pressure-correction equation yields a banded matrix.  
This classical formulation can be naturally mapped onto quantum algorithms; both an initial quantum state and a \(d\)-diagonal matrix representing the linear system are prepared and subsequently solved using techniques such as the quantum singular value transform (QSVT)~\cite{Gilyen2019,Low2019,Martyn2021}.\\

\begin{figure}[!t]
    \centering
    \begin{subfigure}[b]{0.58\textwidth}
        \centering
            \includegraphics[width=\textwidth]{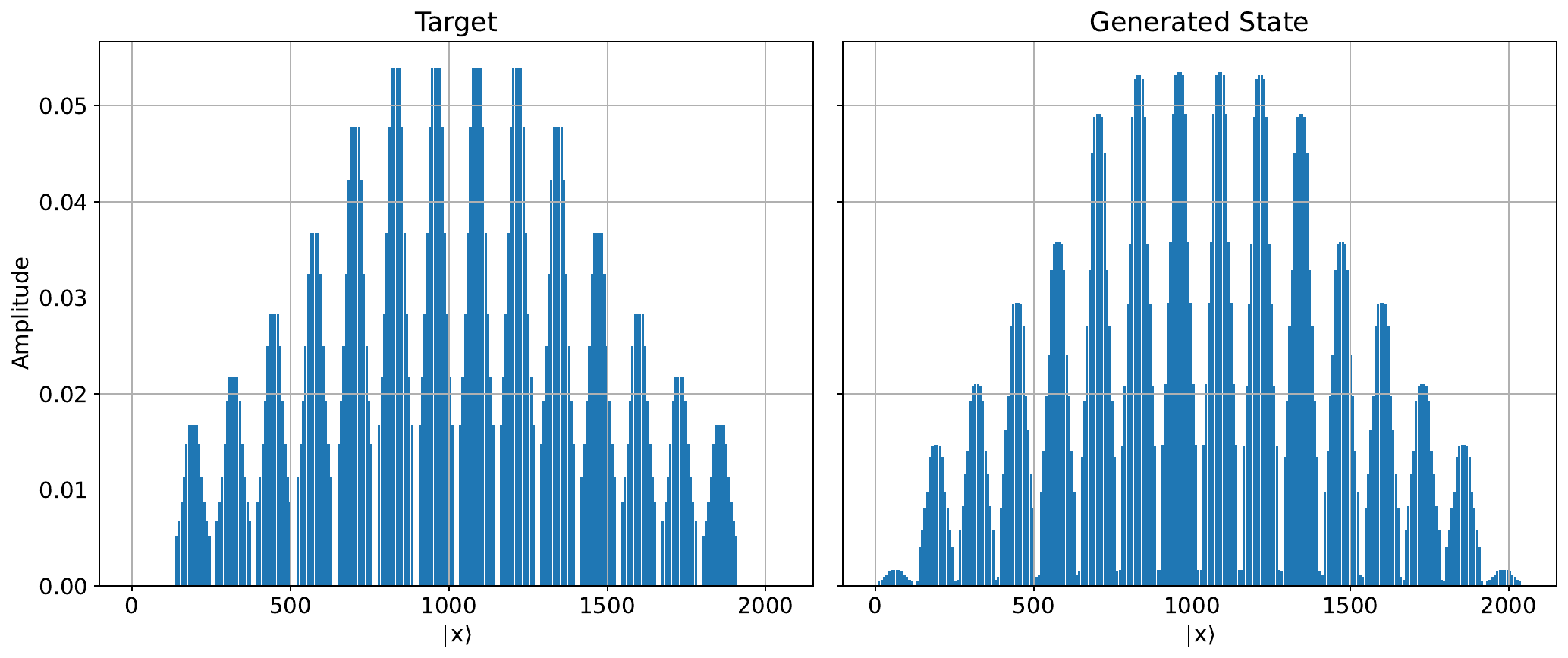}
        \caption{State Reconstruction with MPS. }
        \label{fig:mps_RR}
    \end{subfigure}
    \begin{subfigure}[b]{0.41\textwidth}
        \centering
            \includegraphics[width=\textwidth]{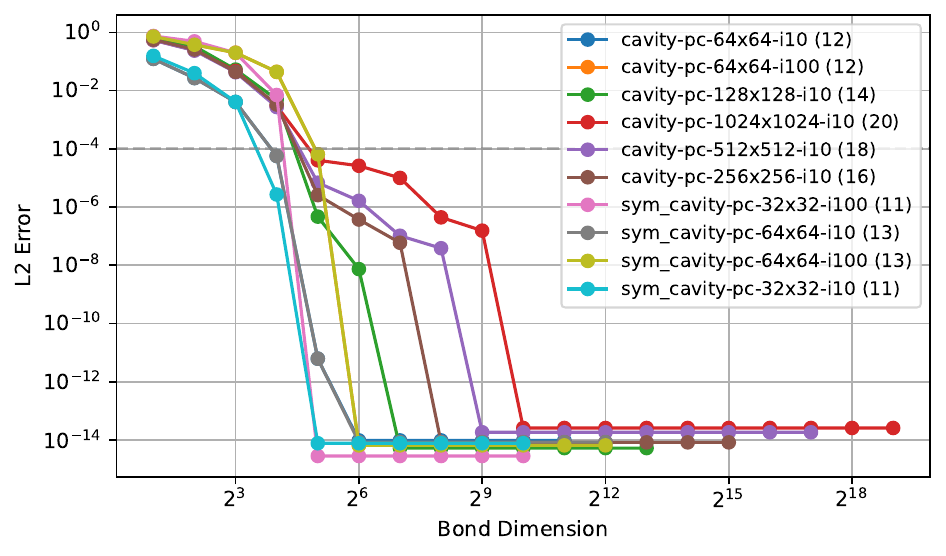}
        \caption{Reconstruction Error Using MPS}
        \label{fig:RR_data}
    \end{subfigure}
    \caption{\textbf{State preparation for 2D-LDC simulations using MPS.} (a) Approximation of the target quantum state using a matrix product state with bond dimension $\chi=2$. (b) Reconstruction error $|||\psi\rangle - |\psi^\prime\rangle||_2$ as a function of the MPS bond dimension $\chi$. The original state $|\psi\rangle$ corresponds to solutions of CFD problems with increasing grid sizes; the effective Hilbert space dimension ranges from $2^{11}$ to $2^{20}$, as indicated in the legend. The reconstructed state $|\psi^\prime\rangle$ is obtained via MPS compression.}
    \label{fig:msp-state-prep-ldc}
\end{figure}

Beginning with the initial state, we applied our framework to instances ranging from $11$ to $20$ qubits, observing a clear tendency toward preparation via matrix product states (MPS). As shown in Fig.~\ref{fig:mps_RR}, for an 11-qubit system and a target accuracy of $\varepsilon = 1\mathrm{e}{-3}$, a bond dimension as small as $2$ suffices to reconstruct the desired state accurately. This is particularly notable, as it enables the recovery of $2^{11}$ amplitudes using only 11 two-qubit gates. During the selection process, a weighted error model was adopted, assigning $70\%$ of the budget to the precision error $\varepsilon_p$ and $30\%$ to the approximation error $\varepsilon_a$, thereby prioritizing high-fidelity state encoding.\\

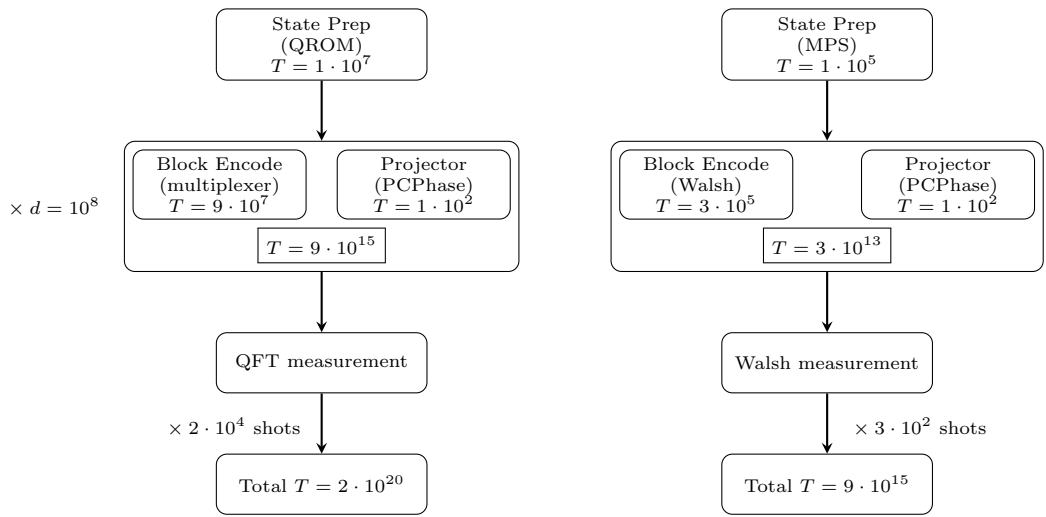
\begin{figure*}[b]
  \centering
  \begin{tikzpicture}[
    block/.style={
      rectangle, draw, rounded corners, minimum width=2.8cm, minimum height=0.8cm,
      font=\scriptsize, align=center
    },
    subblock/.style={
      rectangle, draw, minimum width=2.3cm, minimum height=0.6cm,
      font=\scriptsize, align=center
    },
    arrow/.style={-stealth, thick},
    label/.style={font=\scriptsize},
    scale=1.6
  ]

  \node[block] (prepQ) {State Prep\\(QROM)\\$T=1\cdot10^7$};
  \node[block, below=0.8cm of prepQ] (encQ) {
    \begin{tikzpicture}[every node/.style={subblock}]
      \node (beQ) {Block Encode\\(multiplexer)\\$T=9\cdot10^7$};
      \node[right=0.4cm of beQ] (projQ) {Projector\\(PCPhase)\\$T=1\cdot10^2$};
    \end{tikzpicture}
    \\[2pt] $\boxed{T=9\cdot10^{15}}$
  };
  \node[label, left=0.2cm of encQ.west] {$ {}\times d=10^8$};
  \node[block, below=0.8cm of encQ] (measQ) {QFT measurement};
  \node[label, below=0.2cm of measQ, xshift=-1.2cm] {${}\times2\cdot10^4$ shots};
  \node[block, below=0.8cm of measQ] (totQ) {Total $T=2\cdot10^{20}$};

  \draw[arrow] (prepQ) -- (encQ);
  \draw[arrow] (encQ) -- (measQ);
  \draw[arrow] (measQ) -- (totQ);

  \begin{scope}[xshift=4.2cm]
    \node[block] (prepW) {State Prep\\(MPS)\\$T=1\cdot10^5$};
    \node[block, below=0.8cm of prepW] (encW) {
      \begin{tikzpicture}[every node/.style={subblock}]
        \node (beW) {Block Encode\\(Walsh)\\$T=3\cdot10^5$};
        \node[right=0.9cm of beW] (projW) {Projector\\(PCPhase)\\$T=1\cdot10^2$};
      \end{tikzpicture}
      \\[2pt] $\boxed{T=3\cdot10^{13}}$
    };
    \node[block, below=0.8cm of encW] (measW) {Walsh measurement};
    \node[label, below=0.2cm of measW, xshift=1.2cm] {${}\times3\cdot10^2$ shots};
    \node[block, below=0.8cm of measW] (totW) {Total $T=9\cdot10^{15}$};

    \draw[arrow] (prepW) -- (encW);
    \draw[arrow] (encW) -- (measW);
    \draw[arrow] (measW) -- (totW);
  \end{scope}

  \end{tikzpicture}
  \caption{\textbf{Matrix inversion workflows.} Comparison of $T$ costs for exact (left) and approximate (right) schemes in the matrix inversion workflow.}
  \label{cost}
\end{figure*}

Table~\ref{tab:combined_gate_counts} (column 1) summarizes the CNOT and T-gate costs for MPS in comparison with several alternative techniques. More importantly, this compact MPS-based representation demonstrates robust performance across a wide range of problem sizes. Figure~\ref{fig:RR_data} shows the overall reconstruction error, computed as the $\ell_2$-norm of the difference between the prepared and target state vectors, as a function of the bond dimension for states with $2^{11}$ to $2^{20}$ amplitudes. Notably, in all considered cases, a bond dimension of $\chi = 2^5$ or greater is sufficient to reduce the reconstruction error below $10^{-4}$, illustrating both the scalability and the practical utility of the MPS approach for this application. Subsequently, we load the $d$-diagonal matrices of the system using the algorithm introduced in Sec.~\ref{sec:be}. Taking as an example an $11$-qubit diagonal, we find that an approximation based on Walsh functions, employing the leading $64$ coefficients, suffices to achieve an error below $\varepsilon = 10^{-4}$. The required resources amount to \( 1.5 \times 10^3 \) CNOT gates and \( 1.7 \times 10^2 \) RZ gates. These results demonstrate that the proposed block-encoding technique is a promising approach for CFD applications, with the precision of the approximation illustrated in Fig.~\ref{fig:diagonal_approx}.\\

Once these two main components are defined, we apply matrix inversion through the QSVT algorithm. The primary strategy consists of approximating the function $f(x)=1/x$ with a polynomial over the interval \([ -1, -\frac{1}{\kappa} ] \cup [ \frac{1}{\kappa}, 1 ]\), where $\kappa$ denotes the matrix condition number~\cite{lapworth2022implicit}. Recent work has shown that constructing such a polynomial can now be performed efficiently, and is no longer the computational bottleneck it once represented~\cite{sünderhauf2025matrix}. To analyze the complete workflow, we consider a CFD instance discretized over a $20$-qubit space, requiring the embedding of a \(2^{20} \times 2^{20}\) tridiagonal matrix and the preparation of a $20$-qubit initial state vector. Achieving matrix inversion in this case necessitated the construction of a polynomial of degree $10^8$. Figure~\ref{cost} presents a resource comparison against the previously employed generic approach, which assumed no structural information about the input.\\

An additional innovation in our approach is the use of the Walsh transform for measurement. In systems of this scale, extracting information from the solution state is challenging, as performing full state tomography is prohibitively expensive. Instead, applying transformations that redistribute the amplitude distribution facilitates the measurement process by producing states that are easier to sample.\\ 

\begin{figure}[t]
    \centering
    \includegraphics[width=0.4 \linewidth]{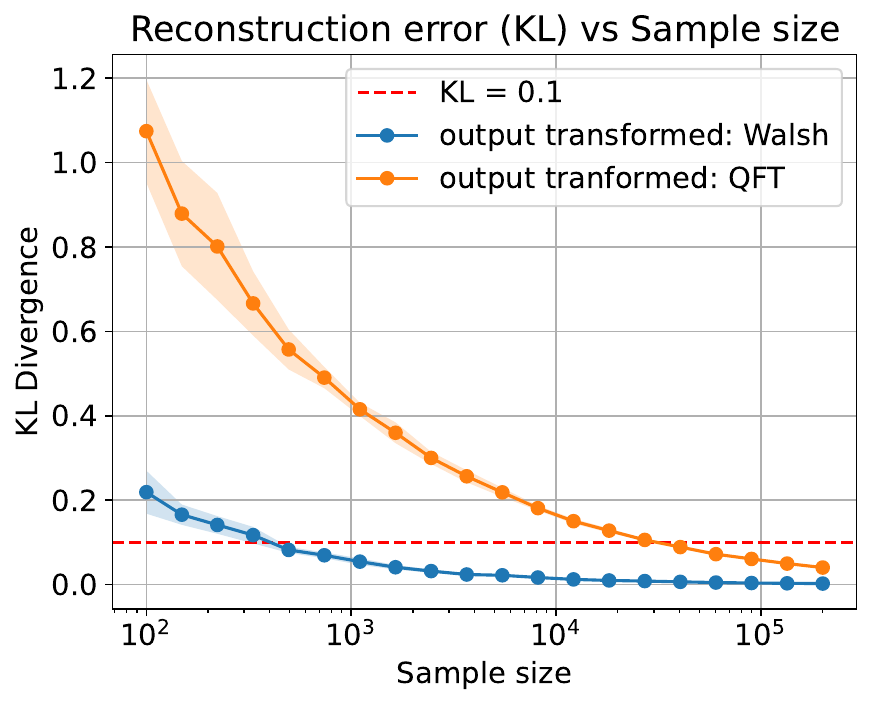}
\caption{\textbf{Reconstruction error versus sample size.} KL divergence measures the distance between probability distributions~\cite{cui2025generalize}, and we assume a divergence tolerance of $0.1$ between the resulting output distributions.}\label{fig:loss}
\end{figure}

Motivated by this, Fig.~\ref{fig:loss} compares the sampling error for different output state transformations as a function of the number of shots, quantified using the Kullback--Leibler (KL) divergence~\cite{cui2025generalize}. A smaller value of \(D_{\mathrm{KL}}\) indicates that $Q$ is closer to $P$, with \(D_\mathrm{KL}=0\) achieved only when $P=Q$ exactly. As shown, the Walsh transform requires significantly fewer shots to achieve lower estimation errors. Remarkably, our complete, structure-aware workflow yields resource savings by a factor of over $\sim$$10^{\;4}$ compared to an exact approach. 

\section{Conclusion}
\label{sec:conclusion}

We have developed two automated workflows, one for state preparation and another for diagonal operator encoding, which together redefine data loading as a structured, systematic stage of algorithm design. Key to this is the integration of PennyLane’s resource estimation framework, which enables evaluation, comparison, and selection of methods on equal footing across a wide portfolio of state-of-the-art algorithms. This integration transforms what was once a heuristic and ad hoc process into a reproducible optimization pipeline, grounding choices in concrete resource costs such as T-gate and CNOT counts. \\

The strength of this approach lies in its ability to turn the vast search space of data-loading algorithms into a navigable design landscape. By explicitly partitioning the error budget into approximation and precision components, the workflows explore whether it is more advantageous to simplify the problem and solve it exactly or to tackle the full problem approximately. This flexibility not only streamlines algorithm selection but also helps uncover strategies that may otherwise be missed. For example, the framework selects matrix product states for fluid dynamics simulations that appear smooth and continuous, problems where Fourier-based methods might have been the intuitive choice. Conversely, for Gaussian states, it identifies Fourier loaders as efficient due to the narrowing of the Gaussian under the Fourier transform. In quantum chemistry, the system recognizes and exploits sparsity in molecular ground states, selecting sparse preparation techniques that scale with the number of dominant amplitudes rather than the full Hilbert space dimension. These results highlight a key principle: efficiency is not found in a universal method, but in recognizing and exploiting the hidden structure of each problem instance.\\

The consequences extend beyond isolated case studies. In the 20-qubit matrix inversion workflow for computational fluid dynamics using quantum singular value transformation, the framework automatically selected an MPS state preparation with modest bond dimension and a Walsh-based diagonal encoding. Coupled with a novel Walsh-based measurement strategy, this led to reductions in resource requirements by more than four orders of magnitude compared to uninformed approaches. In addition to the workflows themselves, we also introduce two new circuit constructions: (i) a more efficient encoding for $d$-diagonal matrices, and (ii) an optimized block encoding for kinetic energy operators, developed and validated through the same systematic methodology. While secondary to the central workflows, these advances highlight the broader value of combining rigorous resource estimation with careful algorithmic design.\\

As quantum hardware continues to mature, with increasingly detailed device-level performance models and noise characteristics, frameworks of this kind will likely be essential. They will serve as the bridge between high-level algorithmic constructions and the low-level circuits that run on physical qubits, ensuring that quantum software stacks evolve in tandem with the capabilities of hardware. By grounding decisions in rigorous benchmarks and by exposing the hidden efficiency in structured problems, such workflows help chart a path toward useful quantum algorithms that are not only theoretically elegant but also practically realizable.

\section*{Code} 
We use \texttt{PennyLane} \cite{2018arXiv181104968B} and its resource estimation functionalities in this work. The complete code is available on the following \href{https://github.com/XanaduAI/QCFDL}{GitHub repository}.

\bibliographystyle{apsrev4-2}
\bibliography{bib}

\clearpage

\begin{appendices}

\renewcommand{\theequation}{\thesection\arabic{equation}}
\renewcommand{\thesection}{\Alph{section}.}

\label{sec:appendix}

\setcounter{equation}{0}
\section{State preparation}

\label{sec:stateprep}

We provide an overview of the key algorithms selected from the literature for the state-preparation branch of the compilation framework, together with the error analysis required to ensure the correct operation of its workflows.

\section*{Multiplexer-based State Preparation}

\begin{figure}[!t]
    \centering
    \begin{subfigure}[b]{0.24\textwidth}
        \centering
        \scalebox{0.8}{
\begin{quantikz}[row sep=0.2cm, column sep=0.3cm]
        \lstick{$\ket{0}$} & \gate[style={fill=blue!10}]{\vec{R}_0} & \ctrl[style={draw,fill=white,shape=rectangle}]{1} & \ctrl[style={draw,fill=white,shape=rectangle}]{2} & \ctrl[style={draw,fill=white,shape=rectangle}]{3} & \qw \\
        \lstick{$\ket{0}$} & \qw & \gate[style={fill=blue!10}]{\vec{R}_1} & \ctrl[style={draw,fill=white,shape=rectangle}]{1} & \ctrl[style={draw,fill=white,shape=rectangle}]{2} & \qw \\
        \lstick{$\ket{0}$} & \qw & \qw & \gate[style={fill=blue!10}]{\vec{R}_2} & \ctrl[style={draw,fill=white,shape=rectangle}]{1} & \qw \\
        \lstick{$\ket{0}$} & \qw & \qw & \qw & \gate[style={fill=blue!10}]{\vec{R}_3} & \qw
    \end{quantikz}}
        \caption{Multiplexed preparation}
        \label{fig:GroverR}
    \end{subfigure}
    \hfill 
    \begin{subfigure}[b]{0.73\textwidth}
        \centering
        \begin{tikzpicture}
        \node (compact) at (0.4,0) {
        \begin{quantikz}[row sep=0.4cm, column sep=0.1cm]
        & \ctrl[style={draw,fill=white,shape=rectangle}]{3} & \qw \\
        & \ctrl[style={draw,fill=white,shape=rectangle}]{2} & \qw \\
        & \ctrl[style={draw,fill=white,shape=rectangle}]{1} & \qw \\
        & \gate[style={fill=blue!10}]{\vec{R}} & \qw
        \end{quantikz}
        };
        \node at (1.0, 0) {$\equiv$};
        \node (expanded) at (6.8, 0) {
        \begin{quantikz}[row sep=0.4cm, column sep=0.15cm]
        & \qw      & \qw      & \qw      & \qw      & \qw      & \qw      & \qw      & \ctrl{3} & \qw      & \qw      & \qw      & \qw      & \qw      & \qw       & \qw & \ctrl{3} & \qw\\
        & \qw      & \qw      & \qw      & \ctrl{2} & \qw      & \qw      & \qw      & \qw      & \qw      & \qw      & \qw      & \ctrl{2} & \qw      & \qw       & \qw      & \qw & \qw\\
        & \qw      & \ctrl{1} & \qw      & \qw      & \qw      & \ctrl{1} & \qw      & \qw      & \qw      & \ctrl{1} & \qw      & \qw      & \qw      & \ctrl{1}       & \qw      & \qw & \qw\\
        & \gate{R_0} & \targ{}  & \gate{R_1} & \targ{}  & \gate{R_2} & \targ{}  & \gate{R_3} & \targ{}  & \gate{R_4} & \targ{}  & \gate{R_5} & \targ{}  & \gate{R_6} & \targ{}       & \gate{R_7} & \targ{}   & \qw
        \end{quantikz}
        };
        \end{tikzpicture}
        \caption{Decomposition of a multiplexed rotation via the M\"ott\"onen method.}
        \label{fig:mottonen_equivalence_compact}
    \end{subfigure}
    
    \vspace{0.5cm} 
    
    \begin{subfigure}[b]{\textwidth}
        \centering
        \begin{tikzpicture}
        \node (compact) at (0,1.36) {
        \begin{quantikz}[row sep=0.33cm, column sep=0.4cm]
        & \ctrl[style={draw,fill=white,shape=rectangle}]{3} & \qw \\
        & \ctrl[style={draw,fill=white,shape=rectangle}]{2} & \qw \\
        & \ctrl[style={draw,fill=white,shape=rectangle}]{1} & \qw \\
        & \gate[style={fill=blue!10}]{\vec{R}} & \qw
        \end{quantikz}
        };
        
        \node at (1.1, 1.36) {$\equiv$};
        \node at (1.6, 1.36) {(i)};
        \node (expanded) at (5.5, 0) {
        \begin{quantikz}[row sep=0.33cm, column sep=0.3cm]
        & \ctrl[style={draw,fill=white,shape=rectangle}]{4} & \qw & \qw & \qw & \ctrl[style={draw,fill=white,shape=rectangle}]{4} & \qw \\
        & \ctrl[style={draw,fill=white,shape=rectangle}]{3} & \qw & \qw & \qw & \ctrl[style={draw,fill=white,shape=rectangle}]{3} & \qw \\
        & \ctrl[style={draw,fill=white,shape=rectangle}]{2} & \qw & \qw & \qw & \ctrl[style={draw,fill=white,shape=rectangle}]{2} & \qw \\
        & \qw & \gate{R_{\frac{\pi}{8}}} & \gate{R_\frac{\pi}{4}} & \gate{R_\frac{\pi}{2}} & \qw & \qw \\
        & \gate[style={minimum height=1.2cm}, wires = 3]{QROM} & \ctrl{-1} & \qw & \qw & \gate[style={minimum height=1.2cm}, wires=3]{QROM^{\dagger}} & \qw \\
        & \ghost{QROM} & \qw & \ctrl{-2} & \qw & \qw       & \ghost{QROM} \\
        & \ghost{QROM} & \qw & \qw        & \ctrl{-3} & \qw & \ghost{QROM}
        \end{quantikz}
        };
        
        \node at (10.2, 1.36) {(ii)};
        
        \node (expanded) at (13.0, -0.25) {
        \begin{quantikz}[row sep=0.33cm, column sep=0.4cm]
        & \ctrl[style={draw,fill=white,shape=rectangle}]{4}  & \qw  & \ctrl[style={draw,fill=white,shape=rectangle}]{4} & \qw \\
        & \ctrl[style={draw,fill=white,shape=rectangle}]{3}  & \qw  & \ctrl[style={draw,fill=white,shape=rectangle}]{3} & \qw \\
        & \ctrl[style={draw,fill=white,shape=rectangle}]{2}  &\qw   & \ctrl[style={draw,fill=white,shape=rectangle}]{2} & \qw \\
        & \qw & \ctrl{1} &  \qw & \qw \\
        & \gate[style={minimum height=1.2cm}, wires = 3]{QROM} & \gate[style={minimum height=1.2cm}, wires=4]{Adder} & \gate[style={minimum height=1.2cm}, wires=3]{QROM^{\dagger}} & \qw \\
        & \ghost{QROM} & \ghost{Adder} & \ctrl{-2}      & \ghost{QROM} \\
        & \ghost{QROM} & \ghost{Adder} & \qw        & \ghost{QROM} \\
        \lstick{$\ket{\text{QFT}_1}$}         & \qw  & \qw     & \ghost{Adder}   & \qw
        \end{quantikz}
        
        };
        
        \end{tikzpicture}
        \caption{Decomposition of a multiplexed rotation via QROMs.}
        \label{fig:qrom-circuit}
    \end{subfigure}
    
    \caption{\textbf{State preparation algorithm with multiplexers.} (a) The representative circuit, where each of the multiplexers corresponds to the $2^m$ multicontrol-RY gates, where $m$ is the number of control qubits. (b) Decomposition of a multiplexed rotation via the M\"ott\"onen method. The angle of each of these rotations is calculated using the Grover–Rudolph algorithm~\cite{grover2002creating}. (c) Decomposition of a multiplexed rotation via QROMs with (i) controlled rotations and (ii) phase gradient register. QROM encodes the binary representation of the angle to be rotated, and the central operator is responsible for converting this binary representation into the corresponding rotation. The state $\ket{\text{QFT}_1}$ represents the phase gradient state that can be reused during the computations.}
    \label{fig:exact-state-prep}
\end{figure}
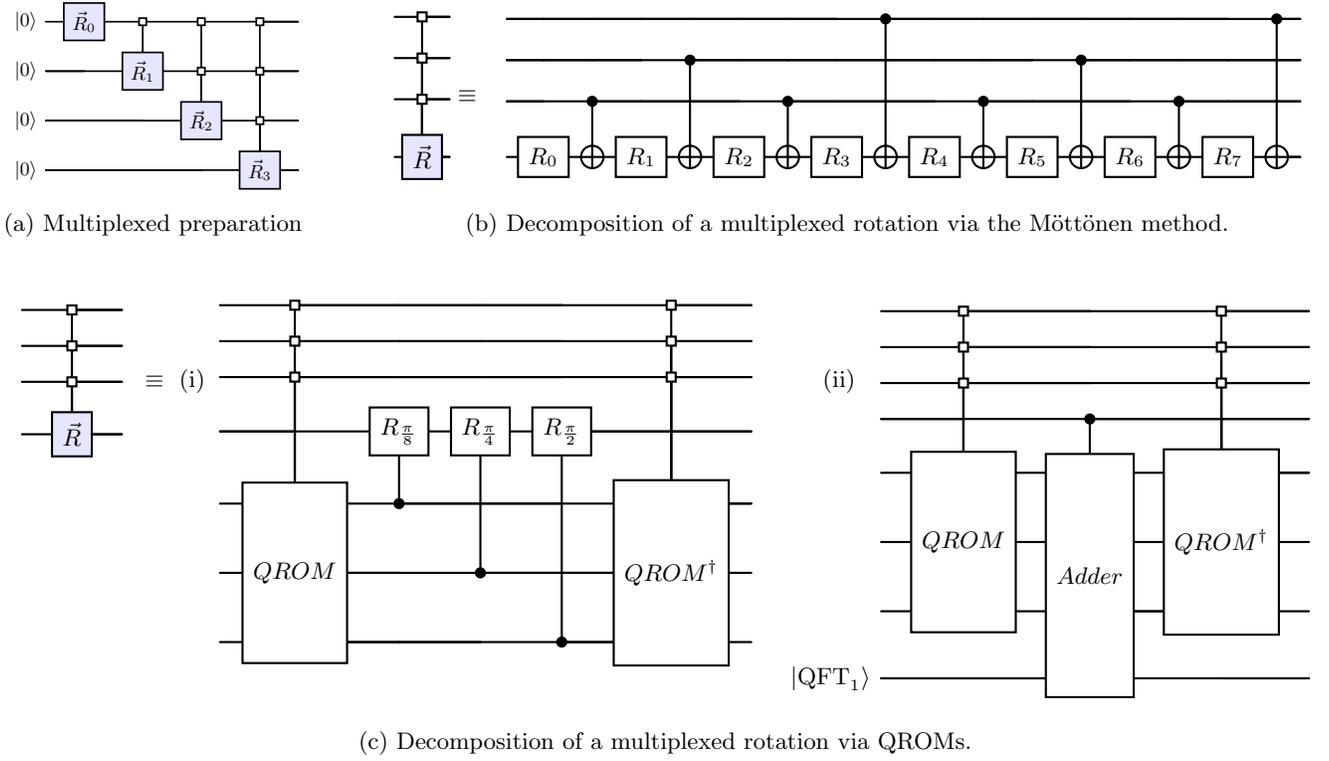

Multiplexer methods are the leading approaches for state preparation without approximation errors \cite{grover2002creating}. These are illustrated in the state preparation circuit presented in Fig.~\ref{fig:GroverR}. The rotation angles $\vec{R}_i$ required to encode the amplitudes of the target vector $\vec{\alpha}$ are computed recursively via the Grover–Rudolph algorithm~\cite{grover2002creating}. The decomposition of these multiplexers depends critically on the availability of auxiliary qubits, as this directly impacts the strategy for decomposing them into native quantum gates. For example, when no such wires are available, the Möttönen method~\cite{mottonen2004transformation} rewrites each multiplexer into a fixed sequence of single-qubit rotations and CNOT gates, as illustrated in Fig.~\ref{fig:mottonen_equivalence_compact}.\\

Conversely, when ancillas are available, a more asymptotically efficient decomposition can be achieved using quantum read-only memory (QROM) techniques based on the \textsc{SelectSwap} framework~\cite{Low_2024}, as shown in Fig.~\ref{fig:qrom-circuit}\,(i). The procedure begins by using a QROM to load the binary representation of the multiplexer rotation angles into $m$ qubits. A sequence of $m$ controlled rotations then converts this binary encoding into the desired rotation, and a second QROM uncomputes the register used to store the angles. This construction can be further optimized by integrating a phase-gradient resource state~\cite{Gidney_2018, O_apos_Brien_2025}, which replaces explicit $R_Z$ rotations with controlled modular additions (Fig.~\ref{fig:qrom-circuit},(ii)). This optimized approach is significant because it eliminates synthesis errors originating from Clifford+T \cite{Kliuchnikov_2023} decomposition, effectively setting the per-gate synthesis error to $\delta_G=0$.\\

Regardless of the chosen decomposition strategy for the multiplexers, obtaining precision requirements for underlying operations is a critical aspect of their operation. In the multiplexer-based state preparation scheme, exactly $2^n - 1$ rotations are required \cite{mottonen2004transformation}, where $n > 0$ is the number of qubits. If the total accumulated precision error must be bounded by $\varepsilon_p$, the uniform synthesis error per gate, $\delta_G$, must satisfy:
\begin{equation}
\label{eq:hyp-mot}
\delta_{G} \leq \frac{\varepsilon_p}{\sqrt{2^n - 1}}.
\end{equation}
Since $\delta_{G}$ directly influences the T-count as $\mathcal{O}(\log_2(1 / \delta_G))$, it becomes a key parameter in resource estimation when decomposing via Möttönen and the QROM method. However, in the phase-gradient QROM case, where this error is eliminated, the only remaining error source is the truncation of representing these $2^n - 1$ rotation angles with a finite number of qubits. By analyzing this truncation, we can derive the number of precision qubits $m$ required to keep the error below $\varepsilon_p$. In order to do that, we assume that $2\pi\theta$ denotes an ideal rotation angle and its truncation $2\pi\theta^{\prime}$ with $m$ binary digits ensures that $|\theta - \theta^{\prime}| < 2^{-m}$. Then, the difference between the corresponding rotations can be bounded as \(\| \mathrm{R_Z}(2\pi\theta) - \mathrm{R_Z}(2\pi\theta^{\prime}) \|_2 < 2^{-m}  \pi \) by using a first-order Taylor series expansion. This allows us to compute \(m\) to determine the minimum resolution of the phase-gradient register required to keep the truncation error below $\varepsilon_p$: 
\begin{equation}
\label{eq:hyp-qrom-sp}
     2^{-m}\pi \cdot \sqrt{2^n - 1} \leq \varepsilon_p \quad\Longrightarrow\quad m \geq \log_2\!\left(\pi \varepsilon_p^{-1}\sqrt{2^n - 1}\right).
\end{equation}

 Ultimately, the number of auxiliary wires is treated as a tunable hyperparameter in our implementation, allowing it to be automatically optimized to trade off qubit overhead against gate complexity.

\subsection*{Sparse State Preparation}

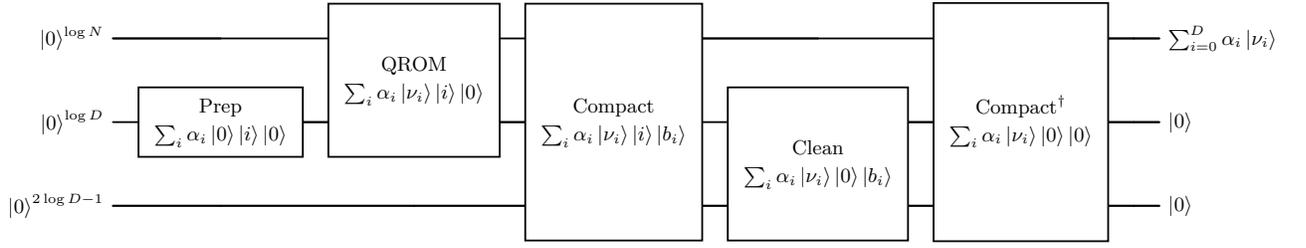
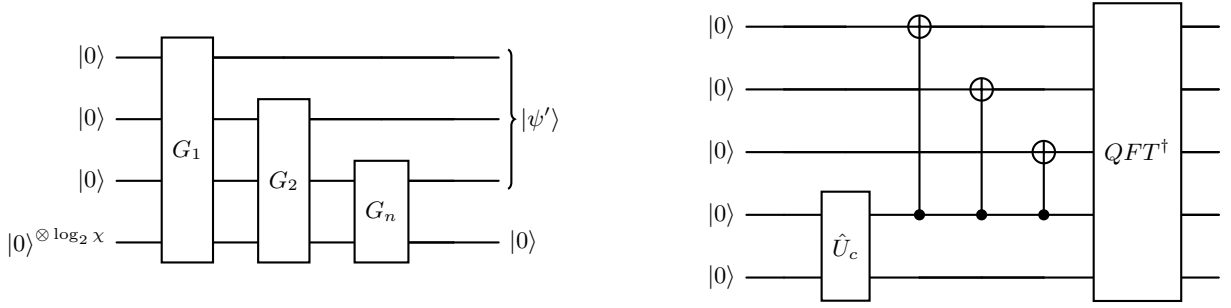
\begin{figure}[!t]
    \centering
    \begin{subfigure}[b]{\textwidth}
        \centering
            \scalebox{0.85}{
            \begin{quantikz}[row sep = 6.0, column sep=0.4cm]
        \lstick{$\ket{0}^{\log N}$} & \qw & \gate[wires = 2]{\begin{array}{c}
              \text{QROM} \\[0.5ex]
              \sum_i \alpha_i \ket{\nu_i}\ket{i} \ket{0}
          \end{array}} & \gate[wires = 3]{\begin{array}{c}
              \text{Compact} \\[0.5ex]
              \sum_i \alpha_i \ket{\nu_i}\ket{i} \ket{b_i}
          \end{array}}  & \qw & \gate[wires = 3]{\begin{array}{c}
              \text{Compact}^{\dagger} \\[0.5ex]
              \sum_i \alpha_i \ket{\nu_i} \ket{0} \ket{0}
          \end{array}}  & \qw & \qw \rstick{$\sum_{i=0}^D \alpha_i \ket{\nu_i}$} \\
        \lstick{$\ket{0}^{\log D}$} 
          & \gate[wires=1]{\begin{array}{c}
              \text{Prep} \\[0.5ex]
              \sum_i \alpha_i \ket{0}\ket{i} \ket{0}
          \end{array}} 
          & \qw 
          & \qw
          & \gate[wires = 2]{\begin{array}{c}
              \text{Clean} \\[0.5ex]
              \sum_i \alpha_i \ket{\nu_i} \ket{0} \ket{b_i}
          \end{array}} 
          & \qw
          & \qw & \qw \rstick{$\ket{0}$} \\
        \lstick{$\ket{0}^{2 \log D -1}$} & \qw & \qw & \qw & \qw & \qw & \qw & \qw \rstick{$\ket{0}$}
        \end{quantikz}
        }
        \caption{Circuit for preparing a sparse quantum state of the form $\sum_i \alpha_i \ket{\nu_i}$, with $i = 0, \dots, D-1$. A QROM first maps each index to its associated basis state. A compact binary identifier is then computed using only CNOT gates~\cite{fomichev2024initial}. Finally, both the index register and the identifier are uncomputed, leaving the desired sparse state.}
        \label{fig:sparse}
    \end{subfigure}
    \vspace{0.5cm}
    \begin{subfigure}[b]{0.49\textwidth}
        \centering
        \begin{quantikz}[row sep={0.82cm,between origins}, column sep=0.6cm]
        \lstick{$\ket{0}$} & \gate[wires=4]{G_1} & \qw                & \qw                & \qw & \rstick[wires=3]{$\ket{\psi'}$} \\
        \lstick{$\ket{0}$} & \ghost{G_1}                              & \gate[wires=3]{G_2}  & \qw                & \qw & \\
        \lstick{$\ket{0}$} & \ghost{G_1}                         &    \ghost{G_2}                  & \gate[wires=3]{G_n}  & \qw & \\
        \lstick{$\ket{0}^{\otimes \log_2 \chi}$}
                          & \ghost{G_1}             & \ghost{G_2}         & \ghost{G_n}         & \qw & \rstick{$\ket{0}$}\\
        \end{quantikz}
        \caption{MPS-based state preparation circuit, where each block $G_i$ acts on one logical qubit and a $\log_2\chi$-qubit ancilla register storing the bond dimension.}
        \label{fig:mps}
    \end{subfigure}
    \hfill
    \begin{subfigure}[b]{0.49\textwidth}
        \centering
        \begin{quantikz}[row sep = 6.0]
        \lstick{$\ket{0}$} & & \qw                   & \targ{}   & \qw       & \qw       & \gate[wires=5]{QFT^\dagger} & \qw \\
        \lstick{$\ket{0}$} & & \qw                   & \qw       & \targ{}   & \qw       & \ghost{QFT^\dagger}         & \qw \\
        \lstick{$\ket{0}$} & & \qw                   & \qw       & \qw       & \targ{}   & \ghost{QFT^\dagger}         & \qw \\
        \lstick{$\ket{0}$} & & \gate[wires = 2]{\hat{U}_c} & \ctrl{-3} & \ctrl{-2} & \ctrl{-1} & \ghost{QFT^\dagger}         & \qw \\
        \lstick{$\ket{0}$} & & &\ghost{U_c}    & \qw   & \qw       &       & \ghost{QFT^\dagger}         
        \end{quantikz}
        \caption{Fourier Series Loader (FSL) circuit, where the unitary $\hat{U}_c$ prepares a truncated set of Fourier coefficients that are transformed to the computational basis via QFT$^\dagger$.}
        \label{fig:fourier}
    \end{subfigure}
    \vspace{-15pt}
    \caption{\textbf{Approximate state preparation methods} using the (a) sparse sum-of-states (SOS) representation, (b) matrix product state (MPS) decomposition, and (c) functional description in Fourier basis.}
    \label{fig:approx-state-prep}
\end{figure}

Another class of algorithms aims to improve preparation efficiency by exploiting specific structural properties of the target state, of which \emph{sparsity} is particularly relevant and can yield substantial resource reductions. The most asymptotically efficient method in this category is the sum-of-states (SOS) algorithm~\cite{fomichev2024initial}, whose cost scales with the number of non-zero amplitudes $D$ rather than the full dimension $2^n$. As shown in Fig.~\ref{fig:sparse}, the algorithm prepares the desired sparse target state $\sum_{i = 1}^D\alpha_i\ket{\nu_i}$ as follows:
\begin{enumerate}
    \item \textbf{Prepare:} Perform a QROM-based preparation to construct $\sum_{i=1}^D \alpha_i \ket{0} \ket{i} \ket{0}$.
    \item \textbf{Map:} Map each $\ket{i}$ to its corresponding determinant $\ket{\nu_i}$ via a second QROM, yielding $\sum_{i=1}^D \alpha_i \ket{\nu_i} \ket{i} \ket{0}$.
    \item \textbf{Compute:} Get a compact binary identifier $\ket{b_i}$ for each $\ket{\nu_i}$, using CNOT gates to obtain $\sum_{i=1}^D \alpha_i \ket{\nu_i} \ket{i} \ket{b_i}$.
    \item \textbf{Uncompute:} Produce the desired SOS state in the system register with all the auxiliary registers reset to $\ket{0}$.
\end{enumerate}

This algorithm can be regarded as an approximation technique. Even when the target state is not sparse, one can truncate the amplitudes by retaining only the $d$ most significant components, zeroing out the rest of the coefficients, thereby introducing a controlled approximation error.

Therefore, a key feature of this method is its explicit balancing of two sources of error: (i) the precision error $\varepsilon_p$ from the QROM-based preparation, and (ii) the approximation error $\varepsilon_a$ from truncating amplitudes. In many cases, amplitudes with small magnitudes can be discarded—retaining only the d most significant terms—without exceeding the global error bound $\varepsilon$, enabling substantial savings in circuit complexity. To manage this trade-off, the sparsity level $d$ and the allocation of the error budget between $\varepsilon_a$ and $\varepsilon_p$ are treated as tunable hyperparameters in our framework. The system automatically optimizes them to minimize total quantum resources while ensuring that the preparation remains within the specified accuracy threshold.

\subsection*{Matrix Product States}

To mitigate the exponential scaling that can affect exact state preparation methods, our framework incorporates approximate techniques that employ compact state representations, trading small, controllable errors for substantial circuit simplifications. Among these, the matrix product state (MPS) formalism is a tool for representing structured quantum states~\cite{berry2024rapid, martin2024comb}. In this approach, the $n$-qubit target state is approximated as $\ket{\psi} \approx \sum_{b_1,\ldots,b_n} A_n^{(b_n)} \cdots A_1^{(b_1)} \ket{b_1\cdots b_n}$ with each tensor $A_k^{(b_k)}$ of dimension $\chi_k \times 2 \times \chi_{k-1}$, where $\chi_k \leq 2^{n/2}$ and the set of \textit{bond dimensions} $\{\chi_k\}$, encodes the entanglement structure of the state \cite{Malz_2024}; larger values enable more correlations at the cost of increased circuit complexity.\\

The preparation circuit, shown in Fig.~\ref{fig:mps}, is built from blocks $G_i$ that implement the following transformation:
\begin{equation}
    G_i = \begin{bmatrix} A_i^{(0)} & * \\[2pt] A_i^{(1)} & * \end{bmatrix} \quad\quad\implies\quad\quad G_i\,\ket{0}_i\ket{\phi}_\chi = \ket{0}_i\,A_i^{(0)}\ket{\phi}_\chi \;+\; \ket{1}_i\,A_i^{(1)}\ket{\phi}_\chi,
\end{equation}
where only the left column is relevant given the fixed input $\ket{0}_i$. The bond dimension $\chi$ controls the approximation error $\varepsilon_a$. To manage this, our framework performs a binary search over $\chi$ values to find the smallest value meeting the target accuracy, which is calculated by contracting MPS tensors and measuring the distance to the target state. The second error source, the precision error $\varepsilon_p$, arises from synthesizing each block $G_i$ into Clifford+T gates using the ZXZ decomposition~\cite{krol2024q, wierichs2025recursive}. The number of rotation gates required to implement the $m$-qubit $G_i$ unitary is $4^{m}$, which is $4\chi_k^2$ rotations per block in terms of the bond dimension. Hence, the total number of rotations required in the preparation would be $4\sum_{k=1}^n \chi_k^2$, and the error-tolerance ($\delta_{G}$) required per rotation must satisfy:
\begin{equation}
   \label{eq:hyp-mps}
   \delta_G < \varepsilon_p \; \bigg[4\sum_{k=1}^n\chi_k^2\bigg]^{-\frac{1}{2}}.
\end{equation}
Therefore, the MPS method introduces two tunable error sources, $\varepsilon_a$ from tensor compression \cite{camaño2025successive} and $\varepsilon_p$ from gate synthesis. Our framework jointly optimizes to minimize resource cost under the global constraint $\varepsilon = \varepsilon_a + \varepsilon_p$.

\subsection*{Fourier Series Loader}

A second approximate state preparation technique implemented in our framework is the \emph{Fourier Series Loader} (FSL)~\cite{Moosa_2023}, which leverages compact functional descriptions of many quantum states. Specifically, it assumes that the amplitude profile of an $n$-qubit target state can be expressed using a smooth or structured function $f : [0,1] \to \mathbb{C}$ as
\(\ket{\psi} = \sum_{i=0}^{2^n - 1} f\!\left(i \cdot 2^{-n}\right) \ket{i}\). When a small number of Fourier coefficients can be accurately approximated $f(\cdot)$, the FSL algorithm can achieve substantial reductions in quantum resource requirements. As shown in Fig.~\ref{fig:fourier}, the algorithm first uses a dedicated unitary $\hat{U}_c$ to encode the truncated set of Fourier coefficients of $f$ into the quantum register. Then, an inverse quantum Fourier transform (QFT$^\dagger$) is applied to transform this frequency-domain encoding into the target amplitudes in the computational basis.\\

This method introduces two distinct sources of error. The \emph{approximation error} $\varepsilon_a$ arises from truncating the Fourier series to a finite number of coefficients. We evaluate $\varepsilon_a$ by computing the Euclidean distance between the target state and the state obtained from the inverse discrete Fourier transform of the truncated coefficients; a procedure that has a classical computational complexity of $\mathcal{O}(2^n n)$ \cite{cooley1965fft}, where $n$ is the number of qubits. For smooth or highly structured functions $f$, the Fourier coefficients decay rapidly, making it possible to achieve small $\varepsilon_a$ with only a few terms.\\

The second error source $\varepsilon_p$ stems from the synthesis of the unitary $U_c$. This operation requires preparing a quantum state whose amplitudes correspond to the Fourier coefficients, and is performed using QROM-based state loading \cite{grover2002creating}. Consequently, it inherits the synthesis and data-loading overheads discussed before on QROM state preparation. Our framework automatically balances $\varepsilon_a$ and $\varepsilon_p$ within the global error budget $\varepsilon = \varepsilon_a + \varepsilon_p$. This allocation is optimized by analyzing $f$ to estimate the convergence of its Fourier expansion, thereby selecting the number of coefficients and synthesis error-tolerance that minimize circuit complexity while respecting the accuracy constraint.

\subsection*{State Preparation via alias sampling}

Our framework also includes the \emph{alias sampling} method~\cite{Babbush_2018}, a technique that prepares a target state by discretizing its amplitudes into equal-magnitude ``fragments'' and coherently redistributing them among the relevant basis states. Instead of directly preparing $\sum_{i=0}^{L-1} \alpha_i \ket{i}$, the method generates a state of the form $\sum_{i=0}^{L-1} \alpha_i\ket{i} \ket{\Lambda_i}$, where the auxiliary register $\ket{\Lambda_i}$ is entangled with $\ket{i}$. By allowing the presence of such entangled registers, one can achieve more efficient preparations; however, this state is not necessarily useful in all scenarios. The key application of this state arises in the context of block encodings of LCUs. 
In this setting, one aims to construct a block encoding of a Hamiltonian $H = \sum_i \alpha_i P_i$,
where each \(P_i\) is a Pauli operator. To achieve this, an operator \(\mathrm{Prep}\) generates the state
$
\mathrm{Prep}\ket{0} =\sum_i \sqrt{\alpha_i}\ket{i}.
$
Subsequently, one applies the operator \(\mathrm{Select}\) (Sel), which maps the state to
$
\sum_i \sqrt{\alpha_i}\ket{i} \otimes P_i$,

followed by \(\mathrm{Prep}^\dagger\). It can then be shown that the block encoding satisfies
\[
(\bra{0}\otimes I) \; \mathrm{Prep}^\dagger \cdot  \, \mathrm{Sel}\cdot \, \mathrm{Prep} (\ket{0}\otimes I) \;=\; \frac{H}{\lambda},
\]

where $\lambda$ is the 1-norm of the Hamiltonian.
If, instead, alias sampling were employed so that 
$
\mathrm{Prep}\ket{0} = \sum_i \sqrt{\alpha_i}\ket{i}\ket{\Lambda_i}$,
the result would remain unaffected, since for any state it holds that 
\(\braket{\Lambda_i | \Lambda_i} = 1\). \\

State preparation via alias sampling starts from an equal superposition over all $L$ basis states and then fragments and rearranges the amplitudes in such a way that, for each basis state \(\ket{i}\), the sum of the associated fragments matches the target amplitude \(\alpha_i\). In other words, the number of amplitude fragments that remain in \(\ket{i}\), together with the sum of all fragments redirected to it, approximates the target value \(\alpha_i\). \\

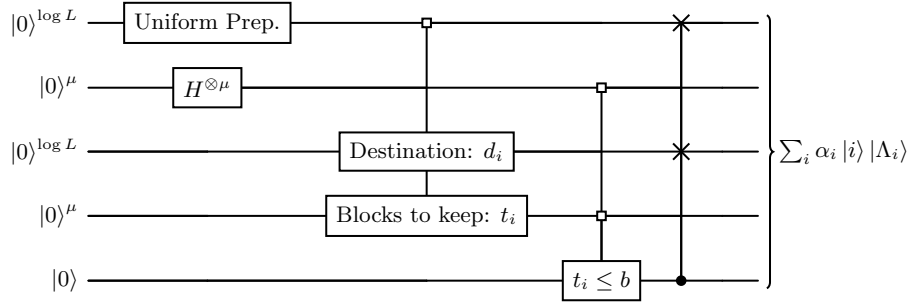
\begin{figure}[t]
    \centering
    \scalebox{0.95}{
        \begin{quantikz}[row sep = 10.0]
            \lstick{$\ket{0}^{\log L}$}
            & \gate[wires=1]{\text{Uniform Prep.}}
            & \ctrl[style={draw,fill=white,shape=rectangle}]{3}
            & \qw
            & \swap{2}
            & \qw
            & \rstick[5]{\rotatebox[origin=c]{0}{$\sum_i \alpha_i \ket{i}\ket{\Lambda_i}$}} \\
            \lstick{$\ket{0}^\mu$}
            & \gate{\text{$H^{\otimes\mu}$}}
            & \qw 
            & \ctrl[style={draw,fill=white,shape=rectangle}]{3}
            & \qw 
            && \\
            \lstick{$\ket{0}^{\log L}$} 
            & \qw 
            & \gate{\text{Destination: $d_i$}}
            & \qw 
            & \swap{2}
            && \\
            \lstick{$\ket{0}^\mu$} 
            & \qw 
            & \gate{\text{Blocks to keep: $t_i$}}
            & \ctrl[style={draw,fill=white,shape=rectangle}]{1} 
            & \qw 
            && \\
            \lstick{$\ket{0}$} 
            & \qw 
            & \qw 
            & \gate{\text{$t_i \leq b$}}  
            & \ctrl{-4}
            &&
        \end{quantikz}
    }
    \caption{\textbf{\textsc{SubPrepare} circuit} for initializing a quantum state with $L$ non-zero amplitudes by discretizing each into $2^\mu$ equal blocks and routing them using QROM-specified thresholds and destinations.}
    \label{fig:subprepare}
\end{figure}

The steps of the method, illustrated in Fig. \ref{fig:subprepare}, proceeds as follows:
\begin{enumerate}
    \item \textbf{Uniform preparation:} Initialize the \emph{source index} register in the equal superposition $\frac{1}{\sqrt{L}} \sum_{i=0}^{L-1} \ket{i} \ket{0^\mu} \ket{0^\ell} \ket{0^\mu}\ket{0}$, where $\mu$ is the number of \emph{block index} qubits, $\ell = \lceil \log_2 L \rceil$ is the size of the state register, and the last qubit is a \emph{flag}.

    \item \textbf{Amplitude fragmentation:} Apply $H^{\otimes \mu}$ to the block index register to generate $2^\mu$ equal-magnitude amplitude fragments for each $\ket{i}$. This yields the state $1/\sqrt{L\,2^\mu} \cdot \sum_{i=0}^{L-1} \sum_{b=0}^{2^\mu-1} \ket{i} \ket{b} \ket{0^\ell} \ket{0^\mu}\ket{0}$, where each block corresponds to a fixed amplitude fragment of size $1/\sqrt{L\,2^\mu}$.

    \item \textbf{QROM lookup:} Query a QROM with register $i$ to retrieve (i) a \emph{threshold} $t_{i} \in \{0,\dots,2^\mu\}$, the number of blocks that remain at $\ket{i}$, and (ii) the \emph{destination} $d_i \in \{0,\dots,L-1\}$, the target indices for the remaining blocks:
    
    $$\frac{1}{\sqrt{L\,2^\mu} }\sum_{i=0}^{L-1} \sum_{b=0}^{2^\mu-1} \ket{i} \ket{b} \ket{d_i} \ket{t_i}\ket{0}.$$
    
    As described in \cite{Babbush_2018}, there exists an algorithm that calculates $d_i$ and $t_i$ efficiently.
    
    \item \textbf{Comparator marking:} Compare the block index $b$ with $t_i$ and set the flag qubit to $\ket{1}$ if $b \geq t_i$. Blocks with the flag set will be routed to $\ket{d_i}$ later. The state generated is:
    \[
        \frac{1}{\sqrt{L}} \sum_{i=0}^{L-1} \left (\frac{1}{\sqrt{t_i}}\sum_{b=0}^{t_i} \ket{i} \ket{b} \ket{d_{i}}\ket{t_{i}} \ket{0} + \frac{1}{\sqrt{2^\mu -t_i}}\sum_{b=t_i+1}^{2^\mu-1} \ket{i} \ket{b} \ket{d_{i}}\ket{t_{i}} \ket{1}\right).
    \]
    \item \textbf{Conditional swap:} Controlled on the flag, swap the source and destination registers, thus moving the marked blocks to their designated targets. The final state is described as:
    \[
        \frac{1}{\sqrt{L}} \sum_{i=0}^{L-1} \left (\frac{1}{\sqrt{t_i}}\sum_{b=0}^{t_i} \ket{i} \ket{b} \ket{d_{i}}\ket{t_{i}} \ket{0} + \frac{1}{\sqrt{2^\mu -t_i}}\sum_{b=t_i+1}^{2^\mu-1} \ket{d_{i}} \ket{b} \ket{i}\ket{t_{i}} \ket{1}\right).
    \]
\end{enumerate}

After these steps, the amplitude associated with each computational basis state $\ket{i}$ is the sum of amplitudes from fragments that stayed plus those that arrived at $\ket{i}$. Formally, the final amplitude $a_i$ can be expressed as
\begin{equation}
    a_i = \frac{1}{\sqrt{L\,2^\mu}}
    \bigg[\quad
        \overbrace{\#\{\,b < t_i \mid b \text{ belongs to } i\,\}}^{\text{blocks that stay}}\quad
        +\quad
        \overbrace{\sum_{\substack{j \neq i \\ d_j = i}} \#\{\,b \ge t_j \mid b \text{ belongs to } j\,\}}^{\text{blocks that arrive}}\quad
    \bigg],
\end{equation}
where the first term counts blocks originating from $\ket{i}$ that remain at the same index (i.e., with block index $b < t_i$), while the second term accounts for the incoming blocks from other states $\ket{j}$ whose assigned destination $d_j$ matches $i$ and whose block index satisfies $b \ge t_j$.
In all cases, each block contributes a fixed amplitude of $1/\sqrt{L\,2^\mu}$. \\

Similar to multiplexer-based methods, this state preparation technique (also known as \textsc{SubPrepare}) introduces no approximation error; the target state is reproduced exactly apart from discretization effects. The only error source is the \emph{precision error} $\varepsilon_p$ from representing each of $L$ amplitudes with a finite number of bits ($\mu$). The minimum value for $\mu$ required is computed as follows: 
\begin{equation}
    \label{eq:hyp-alias}
    \sqrt{L^{-1} 2^{-\mu}} \cdot \sqrt{L} \leq \varepsilon_p \implies  \mu \geq \log_2 \left( \frac{1}{\varepsilon_p} \right).
\end{equation}
Using this, our framework can directly estimate the required number of ancilla qubits and other resources for \textsc{SubPrepare}. This ensures that the method’s resource–error trade-off is optimally balanced for the given hardware and accuracy constraints. \\

\setcounter{equation}{0}
\section{Diagonal Operator Encoding}
\label{sec:diag}

Diagonal encoding \cite{camps2023explicit, zylberman2025efficient} provides an alternative paradigm for quantum data loading, in which classical data is encoded along the diagonal of a unitary operator \(U\), so that \(\langle i | U | i \rangle = \alpha_i\). Unlike state preparation, which constructs a full quantum superposition, diagonal encoding is suited for scenarios where data modulates amplitudes through controlled unitaries or appears as diagonal terms in larger operator constructions. This is common in contexts such as quantum signal processing (QSP) \cite{O_apos_Brien_2025}, polynomial approximation schemes, and block encodings of diagonal matrices \cite{lapworth2024evaluatio}, where the function or matrix of interest is often implemented as a series of controlled phase shifts. \\

A key advantage of this encoding lies in its relaxed normalization constraint: the input vector must satisfy \(\|\vec{\alpha}\|_\infty \leq 1\). This is substantially less restrictive than the $\ell_2$-norm requirement of state preparation \(\|\vec{\alpha}\|_2 \leq 1\), allowing for direct use of unnormalized data from classical simulations or analytic models \cite{O_apos_Brien_2025}. In our analysis, we follow the same methodology used in state preparation, distinguishing between precision error $\varepsilon_p$ and approximation error $\varepsilon_a$ across the different algorithms. An important consideration is that if \( U \) is a diagonal matrix and \( U' \) is its diagonal approximation, then the spectral norm of their difference \( \|U - U'\|_2 \) simplifies to the infinity norm of the vector of their differences \cite{sahasranand2021pnorm}, \( \|\,\text{diag}(U - U')\|_\infty \). Consequently, we adopt the infinity norm of the associated vector as the default norm for this family of loading vectors. \\

We detail the fundamental diagonal encoding algorithms available in the framework, with a focus on their tunable hyperparameters and the associated error trade-offs.

\subsection*{Multiplexer-based diagonal encoding}

In \ref{sec:stateprep}, we discussed how a concatenation of multiplexers can be employed to prepare a specific quantum state. A similar principle underpins \textit{diagonal block encoding}, where a single multiplexer defines a diagonal encoder. Here, the goal is not to create a superposition but to embed an entry \(\alpha_i\) on the diagonal is achieved by setting the angle of the \(i\)th controlled rotation to \(2\arccos(\alpha_i)\), thereby producing a unitary operator with the desired diagonal elements \cite{grover2002creating}. These multiplexers can be decomposed using either the Möttönen method or QROM-based constructions, as illustrated in Figs.~\ref{fig:mottonen_equivalence_compact} and~\ref{fig:qrom-circuit}. A crucial implementation detail is that to ensure that the resulting operator remains diagonal in the computational basis, the rotation gate must act on the first (target) qubit. In contrast, the control qubits are kept in their standard ordering. This configuration guarantees that each computational basis state is correctly addressed and only acquires its corresponding phase, without altering the basis itself. \\

When such an encoder is implemented exactly, i.e., taking \(\varepsilon_a = 0\), these methods require \(2^n\) rotation gates. If we assume that each rotation is synthesized with a consistent error \(\delta_G\), then the condition
\begin{equation}
    \label{eq:hyp-mot2}
    \delta_G < \varepsilon_p \cdot 2^{-n / 2}
\end{equation}
must be satisfied to meet the target precision \(\varepsilon_p\). This stringent tolerance, scaling inversely with the square root of the number of gates, arises from the $\ell_2$ accumulative error set by the framework. As noted previously, the synthesis error \(\delta_G\) related to the number of non-Clifford $T$-gates as $\mathcal{O}(\log_2(1 / \delta_G))$, and therefore plays a central role in resource estimation. This relationship highlights a key trade-off: achieving tighter precision demands more \(T\) gates, thereby increasing the overall cost of the diagonal encoding procedure. Balancing this trade-off between precision and gate complexity is a primary challenge in designing resource-efficient algorithms.

\subsection*{Quantum Signal Processing methods}
\label{sec:QSP}

Another approach for approximating a diagonal operator within an  \(\varepsilon_a\)-tolerance is based on quantum signal processing (QSP)~\cite{mcardle2025quantum, motlagh2024generalized}. The core idea is to treat the diagonal entries as samples of a smooth function \(f:[0,1]\to\mathbb{C}\), evaluated at the discrete points \(x_i = i/N\). QSP then allows us to construct a polynomial transformation that closely matches \(f\) on this grid. We begin by preparing a diagonal Hermitian operator \(V\) (or its block-encoding) whose entries encode a signal. This signal can be embedded in different ways, such as $\langle j|V|j\rangle = \sin\!\left(2\pi j /N\right) \in [-1,1]$. Then, a $d$-degree polynomial \(P\) is selected \cite{mcardle2025quantum}, such that $\big|P(\sin(2\pi x)) - f(x)\big| < \varepsilon_a, \quad \forall\, x \in [0,1]$. Alternatively, if one wishes to approximate \(f(x)\) directly with a polynomial \(P(x)\), rather than with a composition \(P(\sin(2\pi x))\), the signal should be embedded as $\langle j|V|j\rangle = j/N$. Efficient constructions for such block-encodings are readily available and have been employed in prior works \cite{Delgado_2022}.

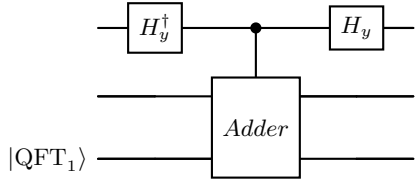
\begin{figure}[t]
\centering
\begin{quantikz}[row sep=0.33cm, column sep=0.4cm]
& \gate{H_y^\dagger} & \ctrl{1} &  \gate{H_y} & \qw \\
& \qw & \gate[style={minimum height=1.2cm}, wires=2]{Adder} & \qw & \qw \\
\lstick{$\ket{\text{QFT}_1}$}         & \qw  & \qw     & \ghost{Adder}   & \qw
\end{quantikz}
\caption{\textbf{Diagonal block encoding of $V$.} Here, $\langle i|V|i \rangle = \sin(2\pi i/N)$ and \(H_y = S H\).}
\label{fig:basicDiag}
\end{figure}

The QSP protocol then implements a unitary whose effective action on the signal register is \(P(V)\), yielding diagonal elements \(\langle i|P(V)|i\rangle \approx f(i/N)\). An efficient circuit for the signal operator \(V\) is depicted in Fig.~\ref{fig:basicDiag}, where \(H_y = S H\) and the auxiliary register is initialized in the phase-gradient \cite{Gidney_2018} state \(|\mathrm{QFT}_0\rangle\). The degree $d$ of the polynomial governs the approximation quality and is treated as a tunable hyperparameter in our framework. Since the QSP sequence involves $\mathcal{O}(d)$ single-qubit rotations, the synthesis error $\delta_G$ of each rotation must satisfy:
\begin{equation}
    \label{eq:hyp-qsp}
    \delta_G \leq \varepsilon_p / \sqrt{d},
\end{equation}
describing the relationship between $d$ and T-counts required for each rotation. This relationship reveals a central trade-off in the overall resource estimation: a higher-degree polynomial $d$ can reduce the approximation error $\varepsilon_a$, but it increases the number of gates and thus the resources required to meet the precision error $\varepsilon_p$.

\begin{figure*}[h]
    \centering
    \begin{subfigure}[b]{.49\linewidth}
    \begin{minipage}{.05\textwidth}
        \caption{}
        \label{fig:diag}
    \end{minipage}%
    \begin{minipage}{0.95\textwidth}
        \begin{quantikz}[column sep=0.4cm]
        \lstick{} & \gate[wires=2]{U} & \qw \\
        \lstick{} &                     & \qw
        \end{quantikz}
        \;\;=\;\;
        \begin{quantikz}[column sep=0.3cm]
        \lstick{}  & \gate{H_y^\dagger} & \octrl{1} & \ctrl{1} & \gate{H_y}  & \qw \\
        \lstick{}  & \qw                & \gate{D}  & \gate{D^\dagger} & \qw  & \qw
        \end{quantikz}
    \end{minipage}
    \end{subfigure}
    \begin{subfigure}[b]{.49\linewidth}
    \begin{minipage}{.1\textwidth}
        \caption{}
        \label{fig:walsh-diagonal-circuit}
    \end{minipage}%
    \begin{minipage}{0.90\textwidth}
        \begin{quantikz}[row sep=0.4cm, column sep=0.45cm]
        \lstick{} & \ctrl{1} & \qw      & \qw      & \qw      & \qw      & \qw      & \ctrl{1} & \qw \\
        \lstick{} & \targ{}  & \ctrl{1} & \qw      & \qw      & \qw      & \ctrl{1} & \targ{}  & \qw \\
        \lstick{} & \qw      & \targ{}  & \ctrl{1} & \qw      & \ctrl{1} & \targ{}  & \qw      & \qw \\
        \lstick{} & \qw      & \qw      & \targ{}  & \gate{R_Z(-2c_k)} & \targ{}  & \qw      & \qw      & \qw
        \end{quantikz}
    \end{minipage}
    \end{subfigure}
    \caption{\textbf{Diagonal block encoding using the Walsh transform.} (a) Equivalent realizations of the diagonal block-encoding, where \(H_y = S H\). The equality on the right removes one control qubit, yielding a more compact construction. (b) Efficient implementation of \(D_{r_k}\) using a CNOT ladder on qubits with \(k_j=1\), followed by a single \(R_Z(-2c_k)\) rotation.}
    \label{fig:diag-walsh-encoding}
\end{figure*}
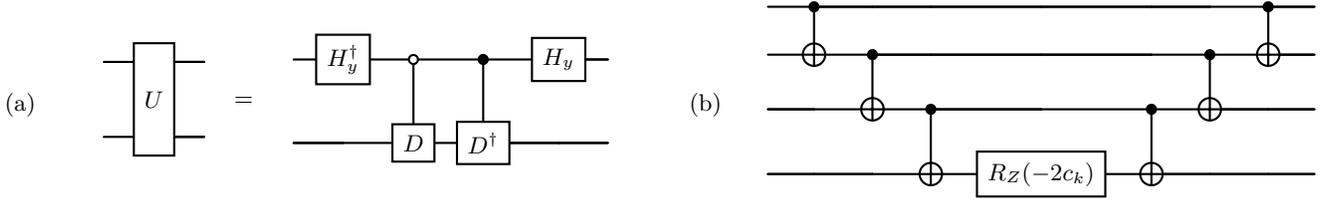

\subsection*{Walsh transform encoding}

Another versatile technique for diagonal block encoding utilizes the Walsh transform, which provides an efficient synthesis of diagonal unitaries from real-valued data. The method can be framed in two main steps:
\begin{enumerate}
    \item \textbf{Reformulate the problem to diagonal unitary synthesis} to obtain the target diagonal unitary $D$ using the phases \(\alpha_j = e^{i\phi_j}\) via $D_{jj} = e^{\, 2i \cdot \arccos(\alpha_j)}$ \cite{zylberman2025efficient}, to encode the desired information in its spectrum. This unitary is then promoted to a block-encoding by controlling it with a single ancilla qubit, as illustrated in Fig.~\ref{fig:diag}. The equivalent circuit representation eliminates one control, resulting in a more compact implementation that reduces the synthesis task to the construction of $D = \sum_{j=0}^{N-1} e^{i g(j/N)} \,\ket{j}\bra{j}$, where \(g:[0,1]\to\mathbb{R}\) is continuous and satisfies \(g(j/N) = 2\arccos(\alpha_j)\).

    \item \textbf{Approximate constructed diagonal unitary in the Walsh basis} by approximating the \(g\) in the Walsh basis. This basis is composed of piecewise-constant orthonormal functions defined by $r_k(x) = \,\mathrm{sign}\!\left[\sin(2^k \pi x)\right] / \sqrt{N}$, yielding a piecewise-constant orthonormal basis with values in \(\{-1/\sqrt{N},1/\sqrt{N}\}\). Expanding \(g\) as $g(x) = \sum_k c_k\, r_k(x)$ produces coefficients \(c_k\) that translate to commuting diagonal operators $D_{r_k}$ = 
    $\sum_j e^{\,i c_k r_k(j/N)}\, \ket{j}\bra{j}$. 
\end{enumerate}

Following~\cite{welch2015synthesis}, we give an efficient circuit-level construction (Fig.~\ref{fig:diag})  to approximate the overall unitary as:
\begin{equation}
    \label{eq:hyp-walsh}
    D \;\approx\; \prod_{k\in\mathcal{K}} D_{r_k} = \prod_{k\in\mathcal{K}} \exp\!\left( i\, c_k \bigotimes_{i} \hat{Z}_i^{k_i} \right),
\end{equation}
where \(k_i\) is the \(i\)-th bit in the binary expansion of \(k\). This structure leads to efficient circuits composed of CNOT ladders and a single \(R_Z(-2c_k)\) rotation (Fig.~\ref{fig:walsh-diagonal-circuit}). In this construction, \(\mathcal{K}\) contains the indices corresponding to the largest coefficients \(c_k\), and if \(g\) is sparse or varies smoothly, the truncation order \(d = |\mathcal{K}|\) will be small, meaning only a subset of Walsh terms is needed. The approximation error \(\varepsilon_a\) is then determined by $d$, where each retained term requires synthesizing one \(R_Z\) rotation to precision \(\delta_G\), which under an incoherent error model should satisfy \(\delta_G < \varepsilon_p \cdot d^{\;-1/2}\) to meet the target precision error ($\varepsilon_p$). The total error is the sum (or, for incoherent errors, the root-sum-square) of the synthesis error and the truncation error from omitting small Walsh coefficients, making $d$ a key hyperparameter for balancing accuracy and gate complexity.

\end{appendices}


\end{document}